\documentclass[12pt]{iopart}

\usepackage{citesort}
\usepackage{aas_macros}
\usepackage{bm}

\expandafter\let\csname equation*\endcsname\relax
\expandafter\let\csname endequation*\endcsname\relax

\usepackage{amsmath,amssymb}
\usepackage{multirow}
\usepackage{threeparttable}
\usepackage{graphicx}

\begin{document}

\title[Machine Learning for Observational Cosmology]{Machine Learning for Observational Cosmology}

\author{Kana Moriwaki$^1$, Takahiro Nishimichi$^{2,3}$, Naoki Yoshida$^{3,4}$}

\address{$^1$Research Center for the Early Universe, The University of Tokyo, 7-3-1 Hongo, Bunkyo, Tokyo 113-0033, Japan}
\address{$^2$Center for Gravitational Physics and Quantum Information, Yukawa Institute for Theoretical Physics, Kyoto University, Kitashirakawa Oiwakecho, Sakyo-ku, Kyoto 606-8502 Japan}
\address{$^3$Kavli Institute for the Physics and Mathematics of the Universe (WPI),
The University of Tokyo Institutes for Advanced Study (UTIAS),
The University of Tokyo, Kashiwa, Chiba 277-8583, Japan}
\address{$^4$Department of Physics, The University of Tokyo, 7-3-1 Hongo, Bunkyo, Tokyo 113-0033, Japan}
\ead{kana.moriwaki@phys.s.u-tokyo.ac.jp, takahiro.nishimichi@yukawa.kyoto-u.ac.jp,naoki.yoshida@ipmu.jp}

\vspace{10pt}
\begin{indented}
\item[]March 2023
\end{indented}

\begin{abstract}
An array of large observational programs using ground-based and space-borne telescopes 
is planned in the next decade. The forthcoming wide-field sky surveys are expected to deliver a sheer
volume of data exceeding an exabyte. Processing the large amount of multiplex astronomical data is technically challenging, and fully
automated technologies based on machine learning and artificial intelligence are urgently needed.
Maximizing scientific returns from the big data requires community-wide efforts.  
We summarize recent progress in machine learning applications in observational cosmology. 
We also address crucial issues in high-performance computing that are
needed for the data processing and statistical analysis.
\end{abstract}

%
%
%
%
%

\section{Cosmology in the big data era}
The last decade witnessed an extremely rapid increase of observational data in astronomy. 
Sky survey is a commonly adopted mode of observation 
that runs a telescope to scan over a wide area of the sky, instead of pointing to specific celestial objects.
Modern imaging devices such as Charge Coupled Devices (CCDs) and Complementary Metal Oxide Semiconductor (CMOS)
sensors can generate a large amount of data in a short time.
For instance, Subaru Hyper-Suprime Cam (HSC) has 104 CCDs on its focal plane, and a single snapshot generates
a billion-pixel image \cite{Miyazaki18}. 
Typically, one-night observation by HSC 
generates a few hundred gigabytes of data.
The Vera C. Rubin observatory LSST Camera
has a greater capability of generating a 3.2 billion-pixel image per one snapshot. As a designated survey telescope, it operates continuously for many years, and is expected to deliver over $500$ petabyte 
of imaging data per year \cite{Ivezic19}.
There are a variety of exciting scientific returns from such 
wide-field, multi-epoch surveys. Discovering
distant supernovae and new types of transient objects, mapping the universe with nearby 
and distant 
galaxies, and probing the nature of mysterious dark matter and dark energy, are noted key scientific cases among many others. All these science goals can be achieved through a sequence of fairly complex data analysis processes.
Efficient data processing is a central issue, but remains technically challenging.

Similar situations can also be found in other research domains,
from life science to engineering, where new experiments and 
sensor technologies boost production and acquisition of
data of impressive quality and quantity. 
Naturally, machine learning (ML) applications have become
increasingly popular in virtually all these research
domains. In astronomy, massive amount of data have already been obtained by ongoing surveys 
such as Dark Energy Survey (DES) \cite{Abbott21}, 
Kilo-Degree Survey (KiDS) \cite{Kuijken19}, 
and Subaru HSC Survey \cite{Aihara22}. 
It is taking over years to produce
major science results after the completion
or occasional data release of each of these observations.
The situation may get even harder with upcoming surveys. Just as an example, the estimated data production rate by Square Kilometre Array (SKA) \cite{Carilli15}
reaches nearly 1 terabyte {\it per second} and a typical 6-hour observation
will produce multi petabytes of data \cite{Scaife20}. Currently no practical 
technology is available to keep storing the
whole data physically, and thus there is urgent need for real-time analysis so that only necessary data
are to be stored. The question is, {\it how do we know and select the necessary data to be stored ?} 
The answer involves the following two aspects; scientific one regarding what are interesting objects and phenomena, and technical one regarding how much data can be processed on-the-fly on available computers. For both the objectives, there is huge demand for efficient and reliable ML or AI-based methods.

In this review, we summarize the rapidly developing research in machine learning applications in observational cosmology.
The topics to be covered are time domain astronomy, cosmology with galaxy surveys, 
and emulation technologies. We introduce
successful applications to real observational 
data, but the limited space does not allow us to provide detailed description of individual ML methods.  To the interested readers, we suggest more technical text books and literature (e.g., \cite{Hastie09,Goodfellow16,Geron17}). There are also a variety of self-learning tools and materials available on the internet.

Throughout this review, we do not distinguish clearly machine learning, deep learning, and artificial intelligence.
We shall refer to the broad range of techniques as ML applications.

\section{Detection and classification of transients}
Transient detection is a classical application of ML. 
Supernovae (SNe) are among the most popular transient objects, which are
known historically since thousands years ago
\cite{Clark82}. A star can end its life by causing a violent and luminous explosion called supernova,
which appears as sudden brightening of a point
(a star) in a galaxy \cite{Branch17}. Its brightness
can even exceed that of the host galaxy, 
and thus it should be straightforward
to spot a supernova if one knows the SN host galaxy 
and compare two images of it taken at different nights.
In Figure 1, we can spot a bright SN at a quick glance.
Doing the same thing for numerous galaxies is a real challenge,
however. A modern large telescope can capture images of
millions of galaxies per night, and
obviously human visual check 
like we do with 
Figure 1 is impractical. 
It is necessary to develop and apply some kind of automated machine 
detection and selection methods.

\begin{figure}
    \centering
    \includegraphics[width=10cm]{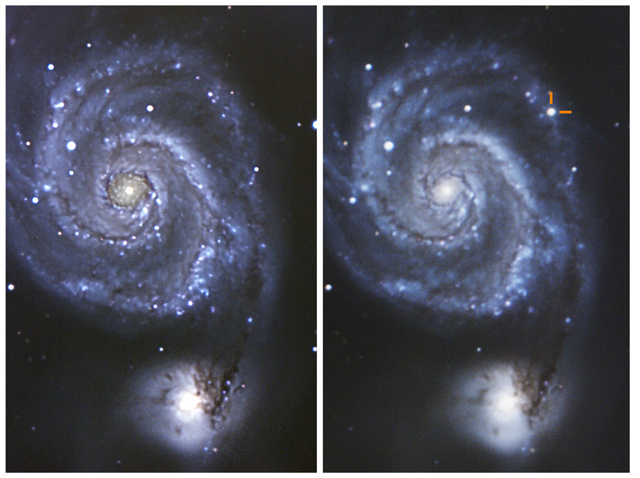}
    \caption{Supernova 2011dh appeared in galaxy M51. The image on the left was taken in 2009, and on the right July 8th, 2011. 
    The supernova is marked by the orange ticks in the upper right
    portion of the right panel. Image credit: Conrad Jung (Chabot Space and Science Center).}
    \label{fig:SN2011dh}
\end{figure}

\subsection{Machine performance}
Before giving an overview of recent development of astronomical transient detection and classification, it would be useful to introduce a few basic diagnostic tools to quantify the performance of ML models.

Let us consider a simple binary classification with positive (real) and negative (bogus) labels. 
Any classification of a sample with a mixture of the two classes yields the following four cases: True Positive (TP), True Negative (TN), False Positive (FP), and False Nagative (FN). 
The number of true positive cases (TP) out of all the positive
cases (TP + FN) yields the true positive rate 
defined as
\begin{align}
\text{TPR} = \frac{{\rm TP}}{{\rm TP+FN}},
\end{align}
whereas the false positive rate is given by
\begin{align}
\text{FPR} = \frac{{\rm FP}}{{\rm FP+TN}}.
\end{align}
The former is also referred to as "recall".
With the same set of quantities, the other often-used metric is
\begin{align}
\text{precision} = \frac{{\rm TP}}{{\rm TP+FP}},
\end{align}
which is also referred to as "purity" for obvious reasons.
The overall accuracy of a classifier can be evaluated by
\begin{align}
\text{accuracy} = \frac{{\rm TP}+{\rm TN}}
{{\rm TP+FP+TN+FN}},
\end{align}
which is the fraction of correct predictions.

Receiver Operation Characteristics (ROC) curves are commonly used to quantify and compare the performance of different methods \cite{Powers11}. There are various manners to represent ROC using different metrics. Often used in astronomy is the one that shows TPR against FPR as in Figure 2.
A classifier can be made flexible to predict the probabilities for each class instead of returning directly a class label. In this case one needs to set a threshold value, say $C_{\rm p}$, so that a probability above $C_{\rm p}$ is considered to be a positive outcome, and otherwise negative. 
Provided with a test sample, the classifier yields an FPR and a TPR for a given $C_{\rm p}$.
By varying $C_{\rm p}$ continuously, one can draw an ROC curve by connecting the resulting set of points of (FPR, TPR).

\begin{figure}
    \centering
    \includegraphics[width=13cm]{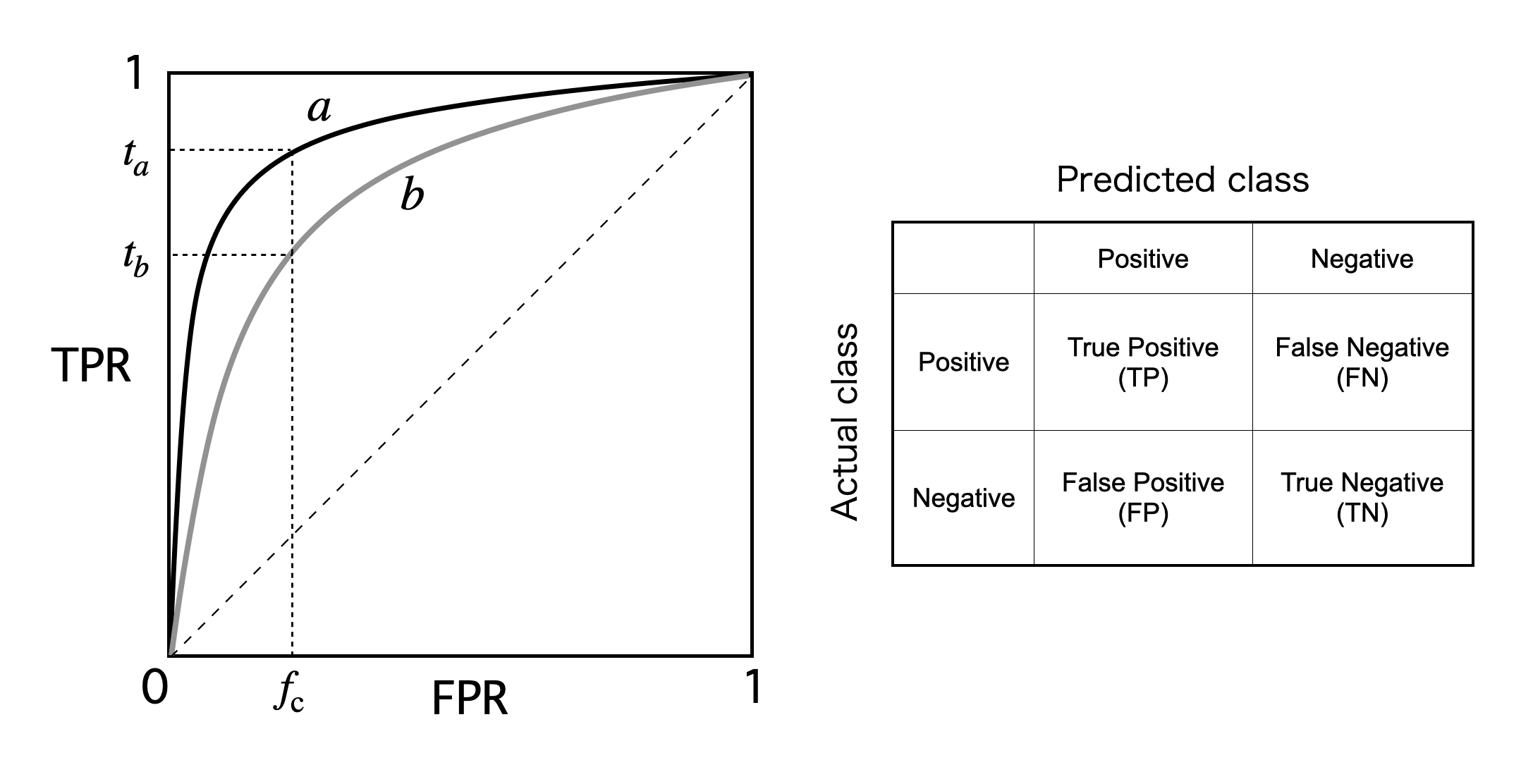}
    \caption{(Left) ROC curves show the overall accuracy of classification methods. Method $a$ yields a larger true positive rate $t_a$, defined in Equation 1, than
    $t_b$ of Method $b$ for a given false positive rate $f_c$ (Equation 2) that is often set as a requirement 
    of an experiment or an observation. 
    (Right) Confusion matrix for a binary classification. Each matrix element is the number of the corresponding case such as True Positive.}
    \label{fig:ROC}
\end{figure}

A good machine classifies accurately and achieves a high TPR while keeping FPR low. 
A perfect classifier would work such that TPR = 1 with FPR = 0,
and then its ROC curve would appear as a
rectangle in the upper left portion. 
Area Under Curve (AUC) measures the integrated area under the TPR curve plotted against FPR. Generally, a large AUC
indicates that the method achieves a 
high TPR with a low FPR. 

Another useful metric is the so-called confusion matrix that summarizes
the classification result for multiple labels (see the right portion of 
Figure 2). Each element in the matrix represents the relative accuracy of the
predicted class compared to the counterparts in the same row or column. Confusion
matrices are especially useful to identify a few particular classes for
which a classifier performs relatively poorly. One can then train the machine 
so that its performance improves for the identified "weak" classes.

The tools introduced here are fairly popular but may not necessarily provide 
clear suggestions when developing a machine for a specific problem.
One may naively think that a machine can be trained so that it achieves a highest score
with a single metric or trained to achieve
as high scores as possible in terms of many different metrics.
In practice, one needs to set practical and empirical requirements, which depend critically on the designated scientific goal. Suppose an astronomical transient survey
is operated so that it can deliver best candidates for Type Ia SNe for follow-up observations. Type Ia SNe can be used
to measure the distance to the galaxies where they occur,
and thus provide a powerful of probe of the cosmic expansion history \cite{Goobar11}.
To search for Type Ia SNe, very accurate identification of the other types,
say Type II SNe, is perhaps unimportant. Instead it is crucial to identify Type Ia 
SNe with high confidence, and without missing scientifically important ones such as
distant Type Ia SNe before their peak brightness.
Clearly, a machine's performance should be judged by using appropriate (combination of) metrics, that are often very specific to the scientific purpose.

\subsection{Transient detection}
In time domain astronomy, detection of transient objects is the first and perhaps the most important step.
Image subtraction yields a number of spots where the local brightness
differs between the two or more images. 
A particular feature
of transient detection and indeed of many other astronomical 
applications is that the fraction of real astronomical phenomena, 
including supernovae and moving objects,
is extremely small compared to other ``bogus" objects
that show up when image subtraction is performed.
The huge disparity of the numbers of
real and bogus objects, which typically amounts to
one out of thousand, demands 
very peculiar tasks to ML applications.
A simple translation of popular ML algorithms would not 
work satisfactorily to the demand, unfortunately.

Bogus pixels are generated in a processed image for a variety 
of reasons. Cosmic rays hit a CCD chip and read-out errors
can occur. Image subtraction should work perfectly if the 
whole image is completely the same except the transient pixels, 
but in practice, two images of the same object taken at
different nights are {\it different} owing to 
many causes
such as sky seeing conditions and miscalibration of telescope pointing (see Ref.~\cite{Starck06} for astronomical image processing).
Fortunately, these typical "errors" can be learnt by a machine
if they are all labeled appropriately. 

Transient detection we discuss in this section
can also be considered as classification of real/bogus objects. 
Applications of an early generation of ML to real/bogus classification
are found in Refs.~\cite{Bailey07,Brink13}.
Automated detection and classification were performed to images collected by Palomar Transient Factory \cite{Bloom12}.
The candidate detection is done by a two-epoch image difference,
and Oarical classifier based on random forest (RF) \cite{Breiman01} is used to distinguish transients from variable stars, active galactic nuclei, and meteorites.
ML algorithms utilize a vector representation of each sample (transient candidate).
The often used vector elements 
are either measured physical quantities such as visual magnitudes or 
may also be some features constructed from
image pixels. The latter is employed
in, for instance, handwritten character recognition \cite{Lecun98} to generate a list of unsupervised features that allow a machine to learn and perform complex tasks. 
Ref.~\cite{Wright15} directly uses image pixel values to construct a feature vector from the data of Pan-STARSS1 Medium Deep Survey,
instead of using transient features that
are defined and derived before the machine selection.
The derived feature data vectors are given to three classifiers (artificial neural network, support vector machine, and RF) to perform real/bogus selection. A combination of the three achieved an
impressively small false negative (missed detection) rate of less than 1 percent.
It is a common practice to adopt
a combination of multiple methods. For example, three methods including a deep NN were adopted to detect
transients in Subaru HSC Survey \cite{Morii16}. The successful application resulted in detection of
65,000 transient objects and 1800 supernovae \cite{Yasuda19}.

Convolutional neural networks (CNNs) have been extensively used
for selecting optical transient candidates
\cite{Gieseke17,Turpin20,Killestein21}.
A CNN-based classifier BRAAI is developed to
separate real astrophysical events from bogus in data from Zwickey Transient Facility \cite{Duev19}. 
BRAAI adopts a custom VGG model \cite{Simonyan14} which consists of sequential layers with activation functions that allow fast learning
with a large imaging data set. 

Most of the popular applications are based on supervised learning.
There are cases where a well trained machine performs poorly
to objects of a particular type. 
A serious problem arises from mislabeling in training data \cite{Ayyar22,Hosenie21}
and also in test data \cite{Northcutt21}, with the latter relatively
less explored even in the general context of image classification.
A practical two-step method has been proposed by Ref.~\cite{Takahashi22}
for the currently operating Tomo-e Gozen Survey 
that detects as many as one million transient candidates per night, which are mostly bogus.

In real observations, classification results by a machine 
or its output "scores" are used as useful metrics, but not as the unique, decisive information. Often many 
other factors are considered in order to select the target objects, and human checks are performed to identify objects for further follow-up observations.
We close this section by discussing a particular need for {\it prompt} detection
of transients. Rapid localization of electromagnetic counterparts of gravitational wave (GW) sources is an important task in time domain astronomy \cite{Levan20}. The angular resolution of LIGO/VIRGO/KAGRA observations,
namely their event localization capability, is $\sim$ 1 square degree as the current best.
This might seem sufficient to point telescopes to the source, but a modern 8-10 meter class telescope is capable of detecting 
thousands of galaxies
within a square-degree field-of-view. One cannot know which galaxy is the host of the GW source. 
Hence prompt identification of the possibly associated optical transient will be of enormous help for rapid follow-up observations,
as has been the case for a binary neutron star merger event GW170817 \cite{Santos17}. 
An urgent task is to detect and identify astronomical transient(s) that is likely
an elecromagnetic counterpart of
the GW event, and quick differentiation from known types will greatly help to generate
a list of targets \cite{Muthukrishna19b,Stachie20}.  
Accumulating a number of such cases will eventually lead to understanding of the physics of the GW events and then to more accurate classification to be performed in the next observational
campaign.

\subsection{Classification of supernovae}
\begin{figure}
    \centering
    \includegraphics[width=15cm]{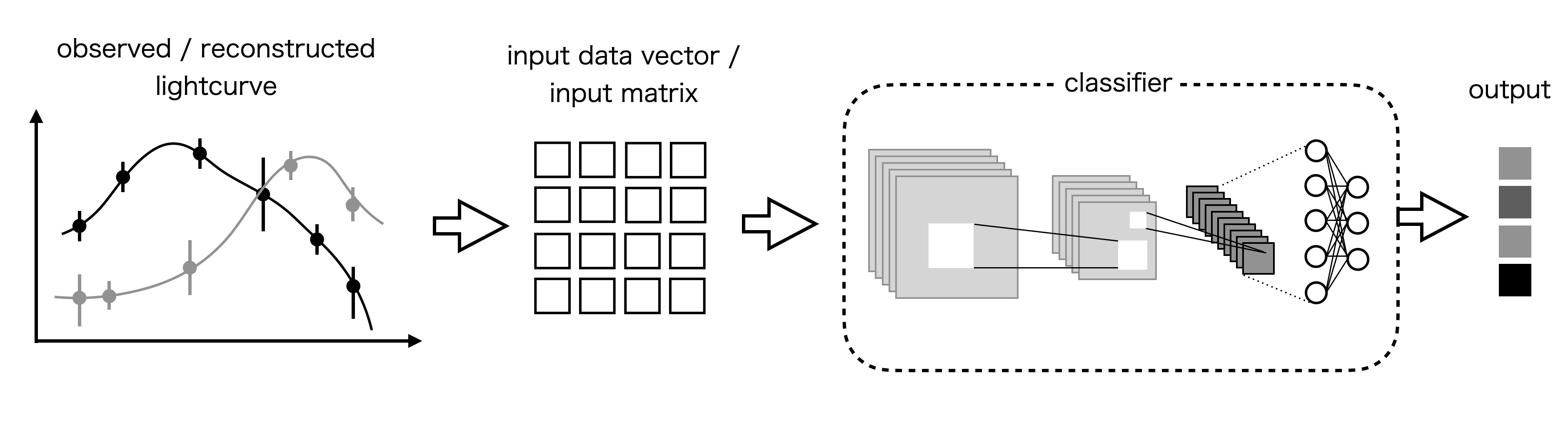}
    \caption{A schematic diagram of photometric supernova classification. The observed lightcurve data are
    directly used or reconstructed by using interpolation techniques such as Gaussian Process. The data vector or matrix are input to the machine classifier, which extracts features of the input data
    and performs classification. In the figure, the classifier
    part is displayed as convolutional neutral networks as an example. The output may be the most likely label for the input, or often the probabilities for multiple labels.
    \label{fig:class_scheme}}
\end{figure}

Classification of supernovae 
(SNe) or of more general transient objects is the next critical
step toward important scientific 
goals such as Type Ia SN cosmology.
Most accurate classification can be done by
spectroscopic observations, which are costly, however, 
and thus can be done only for a limited number of objects. 
Clearly, there is strong demand for {\it photometric} classification of transients, which can be performed with data
of the observed brightness and its time variation of an object
(Figure \ref{fig:class_scheme}).
Rapid photometric 
classification should allow selection of appropriate
targets for prompt follow-up observations.

A straightforward way of classification would be type matching 
using templates of
various types of supernovae \cite{Sako08}.
For an observed brightness variation  (lightcurve) 
as a function of time,
a set of templates are tested to find if any of them describes the observed variation accurately, 
and the best-fit pattern is judged as correct class.
Such templates can be constructed from past observations of well known types of supernovae, 
and/or can be generated from theoretical calculations based on physical modelling of supernovae.

The Supernova Photometric Classification Challenge (SNPCC) is 
aimed at promoting community-wide effort for the development of efficient and accurate
SN classification methods \cite{Kessler10}. The catalogue contains
lightcurves of 5086 Type Ia SNe and 16231 Type II SNe. 
Realistic occurrence rates of the two types are used based upon the observational
estimates of Refs.~\cite{Dilday08,Bazin09}. 
More than 10 groups participated in SNPCC and tested different kinds of algorithms from a nonlinear dimensional reduction technique
\cite{Richards09} to random forest. It would be ideal to compare the classification methods by 
measuring one or a few quantities that characterize
the classification accuracy and the overall efficiency. Unfortunately, SN classification is
a complicated task, and there does not seem to be a convenient single metric. Thus Ref~\cite{Kessler10} 
adopts a practical measure of classification accuracy of Type Ia SNe. Interestingly, even for a single method, the accuracy varies significantly depending on the sample SN redshift, details of the data set, the level of flux errors etc.
It is also shown that each classification method
has its own pros and cons, and there is no clear indication of one particular method performing better than the others. 
The same holds true between conventional
statistical methods and machine-learning based ones.

Since then the SNPCC data set offered a playground for data science in order to test and develop new classification methods.
Statistical analysis methods such as wavelet transformation
and principal component analysis (PCA) have also been applied to extract features from the data,
and several conventional algorithms including Support Vector Machine (SVM) are tested \cite{Lochner16}.
To the SNPCC data, Ref.~\cite{Charnock17} applied deep recurrent neural networks (RNNs), which are suited for analysis of sequential time-series data, and achieved an AUC of 0.986 for a binary 
classification of Type Ia and non-Type Ia.
The RNN-based method is also able to
provide classification probabilities as a function of time. 

The Photometric LSST Astronomical Time-series Classification Challenge (PLAsTiCC) was held in 2018 to promote further development in automated classification of a broad range of transients
\cite{Plasticc18,Hlozek20}. It was hosted in the Kaggle competition platform, and over 1000 teams and individuals participated. 
PLAsTiCC standardized the competition by adopting a 
science-motivated metric. 
For objects $n = 1, ..., N$ that belong
to class $m=1,...,M$, the participants submit
a table of the posterior probability $p(m|d_n)$ 
given the lightcurve data vector $d_n$.
Then the classifier performance is evaluated by 
a weighted log-loss
\begin{equation}
L = - \sum_{m=1}^{M} w_m \sum_{n=1}^{N_m} \tau_{n,m} \, \ln \big[ p(m|d_n)\big],    
\end{equation}
where $\tau_{n,m} = 1$ if the predicted class 
$m_n$ is $m$ (true class) and 0 otherwise.
Larger weights $w_m$ are assigned to rare objects
such as superluminous supernovae so that 
the participants are encouraged to classify accurately the rare classes that otherwise would be overlooked
\cite{Malz19}. This is perhaps close in spirit to 
real science objectives of transient surveys.
Avocado classifier of Ref.~\cite{Boone19} 
marked the highest score with an AUC of 0.957 for classification of Type Ia supernovae. 
It is based on LightGBM, an implementation of gradient boosted decision trees of Ref.~\cite{Ke17}, which is trained with a number of features extracted from a large set of photometric data.
Gaussian Process is adopted to model arbitrary lightcurves in all photometric bands simultaneously \cite{Revsbech18}. 
Clearly, it is the combination of these modern techniques that realized the highest score, rather than a single specific algorithm.

An array of novel methods have been developed and 
have already been applied to real imaging data from HSC \cite{Takahashi20}, 
Pan-STARRS \cite{Villar19}, and DES \cite{Smith20}.
 A high-way architecture which allows fast gradient-based training of deep NNs is used to perform binary and multi-label classification of SNe \cite{Takahashi20}. 
Recurrent autoencoder neural networks (RAENN) is used to achieve a high accuracy of photometric lightcurve classification of the Pan-STARRS Medium Deep Survey Supernovae \cite{Villar20}.
Automatic Learning for the Rapid Classification of Events (ALeRCE) has a two-level structure to allow
hierarchical classification \cite{Sanchez21}. It has been applied to data from Zwicky Transient Facility \cite{Carrasco21} in preparation for LSST. 
Multi-band photometric data are used for a CNN-based classifier of Type Ia SNe with 
only single-epoch observations \cite{Kimura17}. For future transient surveys,
transfer learning is a promising approach to 
enable classification by exploiting information from
learning with data from different telescopes
\cite{Vilalta18}.
 
A new class of image analysis methods have been proposed based on generative models, and extremely large programs for natural language processing are being applied to astronomical data.
A time-series Transformer has been applied to the PLAsTiCC data and has achieved an AUC score of 0.98 \cite{Allam21}.
In Ref. \cite{Qu22}, Gaussian Process (see Section 5 for details) is applied in a two-dimensional domain of time and wavelength, for a CNN to perform photometric classification using early time SN lightcurves.
 
There have also been several efforts in automating spectroscopic classification.
A CNN-based classifier DASH was developed by \cite{Muthukrishna19}.
A total of 17 (sub)types of supernovae
such as Type Ia-91T, Type-Iax, and IIP, IIL are considered.
Existence and identification of subclasses within Type Ia 
is an important topic in supernova cosmology \cite{Sasdelli16}.
Automated spectroscopic classification will be extremely useful
for forthcoming surveys using multi-object spectrograph such as
DESI \cite{DESI16} and Subaru PFS \cite{PFS14}.

 It would be useful to discuss here an important future development for
 optimal observation scheduling, becauese it is closely related to efficient identification of particular types of transients.
 Suppose several transients are detected in their early phase,
when the brightness is rapidly increasing. 
To determine the
type of each transient, its brightness at the next observation 
at some later epoch might be critical. 
The question then is {\it when} one should observe the object
next time and with which wavelength filters, 
in order to determine the
transient type most accurately. 
A Feature-based scheduler is tested for expected cases with LSST \cite{Naghib19}.
In a more general context of telescope operation, a proposal-based scheduler is already implemented for the operation
of LSST \cite{Delgado16}.
Integer programming models have been applied to Zwicky Transient Fascility \cite{Bellm19} to minimize slew times between observations with different sets of filters. It is motivated by the success of
a scheme adopted for the Las Cumbres Observatory Global Telescope Network 
that operate observations over a global network of telescopes \cite{Lampoudi15}.

\subsection{Anomaly detection}
\label{sec:SN-anomaly}
Previous sections consider detection and classification
of transient objects or phenomena that are already known and have been well characterized.
Another extremely important task of an astronomical survey is to make truly new {\it discoveries}.
Ongoing and planned observations will detect numerous objects that are poorly understood or phenomena that are completely unknown so far. 
Detecting and characterizing such {\it outliers} and {\it anomalies}  
will remain an important but non-trivial task.
Anomaly detection is a key technique in many
ML applications, from manufacturing to credit card fraud \cite{Chalapathy19}.

Most of the classification methods introduced in the previous sections utilize 
supervised learning. Given a large set of data, a machine
learns characteristic features of certain types of data or of objects,
and then it is trained to classify a different sample or newly collected data.
This is how supervised learning works usually.
Contrastingly, unsupervised learning is suited for finding
outliers and anomalies, or even
"unknown unknowns" in the data. 
A machine can learn some certain patterns of complex, high-dimensional data
by means of grouping, clustering,
and/or dimensionality reduction
without using prior information on the data sample. 
Without any explicit instruction nor expert labelling, 
it becomes able to identify, in one way or another, outliers that are 
clearly different from known types or are well separated from other major groups of data.
Unsupervised ML has been applied to astronomical transients \cite{Baron19}, and sophisticated
methods have already been developed.
Ref.~\cite{Muthukrishna21} combines a probabilistic deep NN model and
a Bayesian approach to identify rare
transients such as kilonovae and tidal disruption events, whose lightcurves are distinct from those of supernovae.
Models based on unsupervised learning are also able to predict future fluxes from time-series data together with the associated uncertainties.
Ref.~\cite{Villar21} develop 
and test an unsupervised method based on a variational recurrent autoencoder NN (VRAENN)
to deal with unlabeled lightcurve data. A portion of the PLAsTiCC data were used for the training. The VRAENN architecture does not require physical models of transients, and can work with unevenly sampled lightcurve data which 
are converted to encode vectors.
Then an isolation forest with a number of decision trees is used to evaluate an anomaly score for each lightcurve.
Several peculiar SNe including super-luminous and pair-instability SNe were successfully detected by the method.

Anomaly detection algorithms have been applied to various astronomical data sets such as stellar spectra, galaxy images, and photometric lightcurves.
For a specific purpose, Ref.~\cite{Chan22} uses a convolutional variational autoencoder to learn a low-dimensional latent representation of periodic variables. The machine successfully identify anomalous objects in the ZTF data,
which are likely asymptotic giant branch stars or young stellar objects.

\section{Galaxy population and evolution}
\label{sec:galaxy}

Future wide-field surveys 
such as LSST \cite{LSST09}, 
Euclid \cite{Euclid13}, 
and the Nancy Grace Roman Space Telescope \cite{Spergel15} 
will detect billions of galaxies, both
near and far, and their statistical properties as galaxy {\it populations}
will provide rich information on galaxy formation and evolution
over cosmic time.
Also the summary statistics of the large-scale galaxy distributions 
are sensitive probes of cosmology \cite{Mo10}.
Reproducing the observed statistical properties of galaxies
is an important goal of theoretical study of galaxy formation.
Here, we introduce ML applications that are aimed at either 
extracting information from observed galaxy populations
or 
at modeling galaxy formation and evolution.

\subsection{Information extraction from observed data}

Fast and automated measurement of galaxy properties 
is urgently needed to analyze the extremely large data from the future wide-field surveys. 
Automated morphological classification such as distinguishing elliptical/spiral galaxies is one of the most important tasks, and
ML-based classification methods have been developed successfully
thanks to existing large "pre-labeled" datasets such as Galaxy Zoo \cite{Lintott11,Willett13}. 
Photometric redshift estimation is also a crucial task for galaxy surveys, and a wide range of ML applications have been explored
(see Refs.~\cite{Brescia21,Newman22} for
recent development).
ML can also be used to identify specific galaxy images such as strong gravitational lensing effect \cite{Hezaveh17} 
and galaxy merger remnants \cite{Bottrell22},
to detect anomalous objects \cite{Lochner22},
to deblend multiple objects \cite{Boucaud20,Arcelin21},
and to deconvolve point spread functions \cite{Wang22}.
Various ML techniques have been applied for these purposes, and extensive comparisons of the methods have been done \cite{Cavuoti12,Tanaka18,Sevilla-Noarbe18,Rafieferantsoa18,Metcalf19,Henghes21}.
In this section, we will not describe the full details of individual methods, 
but rather focus on the recent progress and general aspects of these ML applications.

A crucial advantage of ML methods is
that one can easily treat or incorporate imaging data in addition to photometric data
and spectral energy distribution (SED).
CNNs are known to be promising ML methods to analyze galaxy images.
Well-trained, sophisticated CNNs are able to classify photometric images as accurately as human experts (professional astronomers) and perform much faster \cite{Huertas-Company15,Cheng22}. 
In strong lens searches, 
many promising lens candidates have been newly discovered with the help of CNNs
from KiDS \cite{Petrillo17,Petrillo19a,He20_KiDS_SL,Li21_KiDS_SL},
the Pan-STARRS 3$\pi$ survey \cite{Canameras21},
and the DESI Legacy Imaging Surveys \cite{Huang20,Huang21,Storfer22}. 
Morphological information is also used in photometric redshift estimation. Refs.~\cite{Pasquet19_photoz,Henghes21_CNN} show that CNN-based frameworks estimate photometric redshifts of SDSS galaxies with higher precision than traditional models that use integrated photometric information alone.
Recently, an innovative image analysis model called Vision Transformer was proposed, which works as efficient and accurate as CNNs in estimating galaxy properties, especially when a large training dataset is available \cite{Lin21,Thuruthipilly22}.
Vision Transformer is considered to be more suited for capturing correlations between distant pixels in an image than CNN, and thus further development for applications in astronomy is to be explored.

In order to train supervised ML models that perform classification and regression tasks, one needs to prepare a large number of labeled training data, either observational 
or computer generated. 
While a supervised machine often outperforms conventional methods when sufficient training data is available, 
it may not be able to maximize its full potential otherwise \cite{Cavuoti12}.
An important technique to compensate for insufficient training data is data augmentation. One can use either analytical methods \cite{Hoyle15} or ML generative models \cite{Ravanbakhsh16} to increase effectively the number of training data.
Another solution is to use unsupervised learning models,
which can work even when little or no labeled data are available.
An illustrative example is morphological classification 
of galaxy images without labels such as elliptical and spiral. 
A machine first divides the images into several groups by a clustering algorithm, and then
the members of each group are visually classified to label the group \cite{Martin20}.
Similar unsupervised models can be applied to regression problems \cite{Geach12}. 
Unsupervised ML techniques are also used for anomaly detection (Sec. \ref{sec:SN-anomaly} in this review). Artefacts such as objects with spurious photometry can be removed in performing statistical analysis, and interesting rare objects can be detected with clustering methods \cite{Solarz17} or generative models \cite{Storey-Fisher21}.

Once a machine is trained and optimized, it can quickly classify or analyze a large set of observed galaxy images and generate a labeled catalogue.
Recently, Ref. \cite{Cheng22} provided one of the largest morphological classification catalogs of more than 20 million galaxies from the DES Year 3 data, with an estimated accuracy of over 99 percent for bright galaxies,
by training a machine with a few thousand galaxies from DES Year 1 data.
Photometric redshift catalogs of 39 million KiDS galaxies \cite{deJong17,Bilicki18},
34 million HSC galaxies \cite{Schuldt21},
a billion DESI Legacy Imaging Survey galaxies \cite{Duncan22},
and 3 billion Pan-STARRS1 galaxies \cite{Beck21} 
have also been recently generated with ML analysis.
The ML models can be applied to the future large observations, which will provide catalogs of billions of galaxies,
and will allow us to study statistical properties and to properly extract cosmological information from the data.
In preparation for this, detailed comparisons between various ML models and other conventional methods are being made for upcoming surveys 
such as LSST \cite{Schmidt20} and Euclid \cite{Euclid20,Euclid22}.

\subsection{Modeling galaxy formation and evolution}
\label{sec:galaxy_model}
A number of numerical simulations have been performed to follow the
formation of galaxies starting from realistic cosmological initial conditions.
Modern simulations follow hydrodynamics as well as gravitational interaction of baryons and DM with incorporating sub-grid models
of star formation and stellar feedback effects.
We refer the readers to recent comprehensive review articles on the computer models of galaxy formation and evolution \cite{Vogelsberger20}.
Such sophisticated computer models provide quantitative predictions for statistics of galaxy populations and their spatial distributions, 
and can also be used to estimate various systematic and statistical errors in 
analyses of data from wide-field surveys.

Properly calibrated numerical models with respect to direct observations are indispensable,
but direct simulations are computationally expensive, and
thus are not well suited to generate a large number of realizations with a sufficiently large volume and high resolution.
There are alternative methods such as semi-analytical models (SAMs) that combine physical models of galaxy formation either with analytic prescription of dark halo formation or with outputs of cosmological $N$-body simulations \cite{Somerville15,Behroozi19}. 
Commonly used DM-only simulations require significantly less computational cost 
than hydrodynamics simulations, allowing one to explore
a large number of physical models that connect the properties of DM halos to those of galaxies.
The essential goal is to populate galaxies on top of DM halos 
or on the DM density field, using multiplex relations derived from  direct observations or from numerical models that are calibrated by observations.

\begin{figure}
    \centering
    \includegraphics[width=10cm]{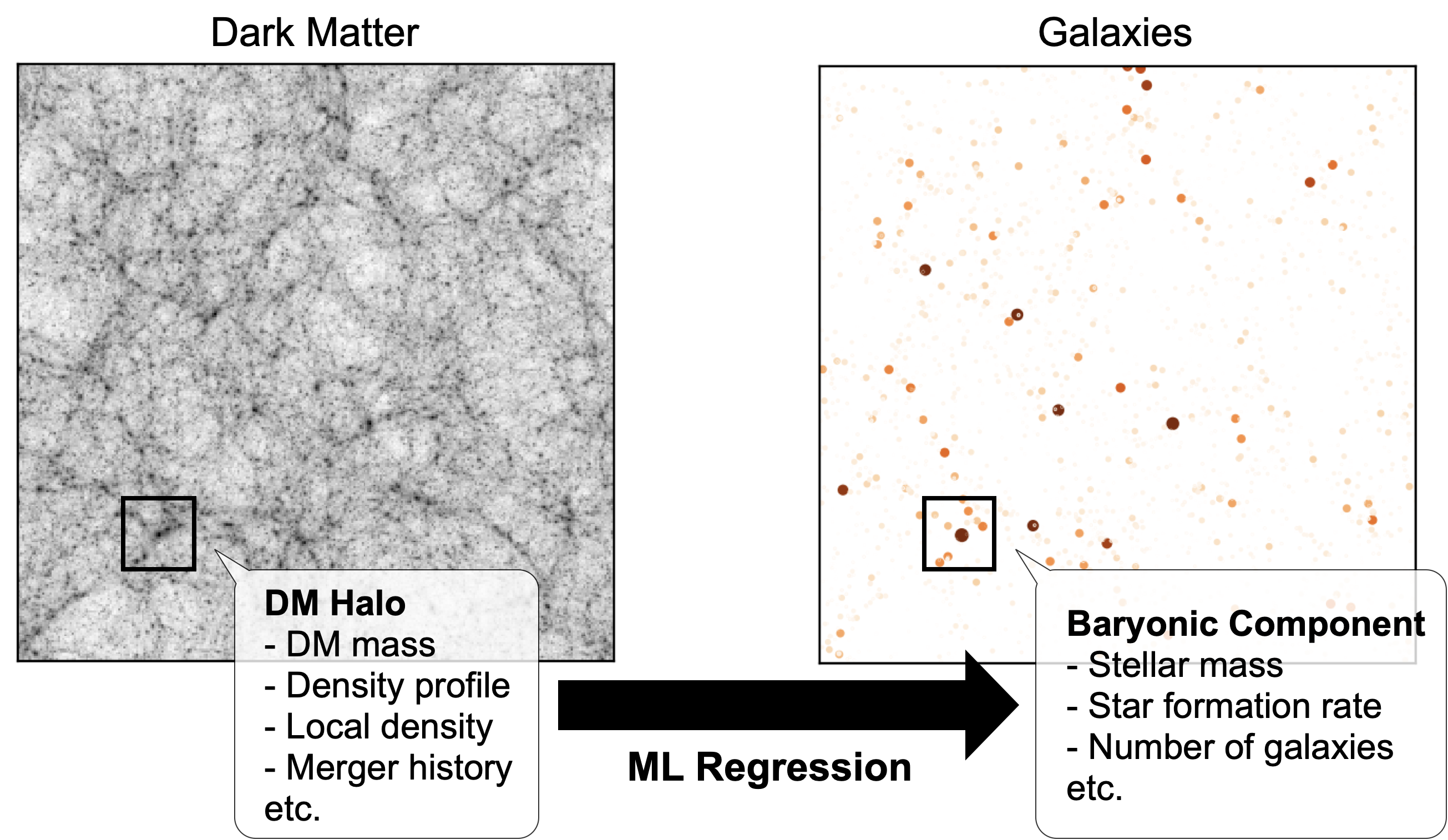}
    \caption{Machine learning can be used to generate a mock galaxy catalog (right) from a DM-only simulation (left). One can use regression models such as decision trees and NNs to infer nonlinear relations between galaxies and DM halo properties.}
    \label{fig:dm-galaxy}
\end{figure}

ML can be used to infer nonlinear relations between galaxy properties
such as stellar mass and star formation rate to the physical properties of the host dark halos and the environment such
as halo mass, density profile, and the local density (Figure \ref{fig:dm-galaxy}).
Various ML models have been proposed to solve this regression problem. 
Earliest studies built regression machines based on well-known ML classifiers such as SVMs and k-nearest neighbor algorithms (kNNs) to predict DM-galaxy connections \cite{Xu13,Ntampaka15}.
Later, advanced techniques such as decision trees, neural networks, and sparse regression models
have been tested and found to perform better \cite{Kamdar16,Agarwal18}.
Models based on decision trees and sparse regression can also quantify the relative importance of halo parameters in determining galaxy properties from training data with diverse quantities \cite{Kamdar16,Morice-Atkinson18,Moster21,Icaza-Lizaola21,deSanti22}.
Several hybrid approaches combining ML and SAM methods are also proposed \cite{Hearin20,Moews21}.
Recent studies extend the ML methods by training a machine with providing more information. For instance, temporal evolution can be implemented by
either feeding information of halo merger trees \cite{Jo19,Jespersen22} 
or using RNNs to analyze sequential snapshot data \cite{Chittenden22}.
Interestingly, models trained with hydrodynamics simulations can perform better than popular SAMs \cite{Jo19}.
An even faster approach is proposed by Refs. \cite{Zhang19,Kasmanoff20},
which can work with a linear DM density distribution as a basis instead of a DM halo catalog. 
The authors use three-dimensional CNNs to generate galaxy distributions 
from the underlying DM density distribution.
    Such models indeed capture the features of the spatial distribution of galaxies, 
    but also reproduce low-order statistics such as the density power spectra with high accuracy.

Obviously the accuracy of a ML method critically relies on the accuracy of the training numerical/theoretical models. 
Unfortunately there are more than a handful of features of real galaxies that are {\it not} reproduced 
by popular numerical models.
Thus it is important to reduce the generalization error of ML owing to 
limited or "biased" training data,
and to understand which physical models are more accurate.
Recently, CAMELs project \cite{Paco21,CMD22} provided a large set of numerical simulation data
that are mostly publicly available outputs of cosmological simulations of galaxy formation.
The available simulations differ in physics modelling, assumed cosmology, and numerical methods. Fortunately, the different sets of data can serve as training data 
with model "variations". The rich dataset can be used to robustly predict the connection between DM haloes and galaxies 
\cite{Villanueva-Domingo22,Shao22}
and to constrain cosmological parameters from observations \cite{Villaescusa-Navarro21b} without being biased by a specific model.
The training data covering a sufficiently broad range of physical models also enable us to explore the most appropriate parameters of the subgrid models in the simulation by making direct comparison of simulated galaxies with the observed ones \cite{Maccio22,Thiele22}.

Combined with low-cost DM-only simulations, the above ML techniques allow us to generate a number of mock galaxy catalogs encompassing a large cosmological volume, 
which are difficult to produce with current hydrodynamics simulations.
The predicted clustering features of galaxies are of great importance in 
the analyses of data from future galaxy surveys.

\section{The large-scale distribution of galaxies and dark matter}
\label{sec:LSS}

Virtually all the ongoing and future cosmology surveys are aimed at probing the large-scale distribution of matter and galaxies in various manners.
In this section, we introduce three major methods and the related 
ML applications. There are different kinds of needs and technical issues in ML for the different observational methods. 

Generative models are becoming increasingly popular in many research areas from image analysis to natural language processing.
In cosmology, applications can be found, for instance, in generating fine images of galaxies and in
reconstruction of large-scale dark matter distribution. 
In this section, we introduce recent development and discuss promising uses of generative models, in addition to ML parameter inference techniques.

\subsection{Dark matter distribution probed with gravitational lensing}

Weak lensing (WL) has been established as an essential
cosmological probe of matter distribution in and around galaxies
and galaxy clusters \cite{Mandelbaum18}. Conventionally, summary statistics of lensing shear and convergence fields
are used to characterize the matter distribution and to infer cosmological parameters such as matter density $\Omega_{\rm m}$ and the density fluctuation amplitude $\sigma_8$ (see Section 5 in this review). 
There have been several proposals to use reconstructed two-dimensional \cite{Lu22} or three-dimensional density fields \cite{Lazanu21} to
study baryonic effects as well as to determine cosmological parameters from density distribution features that are not well captured by conventional summary statistics \cite{Ribli19a}. 

Convolutional neural networks (CNNs) are used for cosmological regression analysis using noisy convergence maps \cite{Fluri18,Ribli19b}. 
DeepMass of Ref.~\cite{Jeffrey20} is based on the U-Net architecture
that is originally developed for biomedical image segmentation
\cite{Ronneberger15}. The U-Net structure allows
to learn many features from a large training data set.
It is applied to reconstruct the
mass distribution from mock Dark Energy Survey science verification data,
to show better performance than Wiener filtering in terms of mean square error. 
Several studies show that CNNs outperform conventional parameter inference methods that utilize low-order summary statistics such as convergence power spectrum and two-point correlation functions \cite{Gupta18}.
CNNs are considered to have a particularly powerful
ability of capturing complex features in two-dimensional images, beyond simple summary statistics, 
and thus are able to distinguish subtle differences in 
the matter distributions in different cosmological models.
Saliency methods \cite{Simonyan13_saliency}, which evaluates the importance of individual input pixels, are used to interpret the function of neural networks \cite{Matilla20}.

\begin{figure}
    \centering
    \includegraphics[width=12cm]{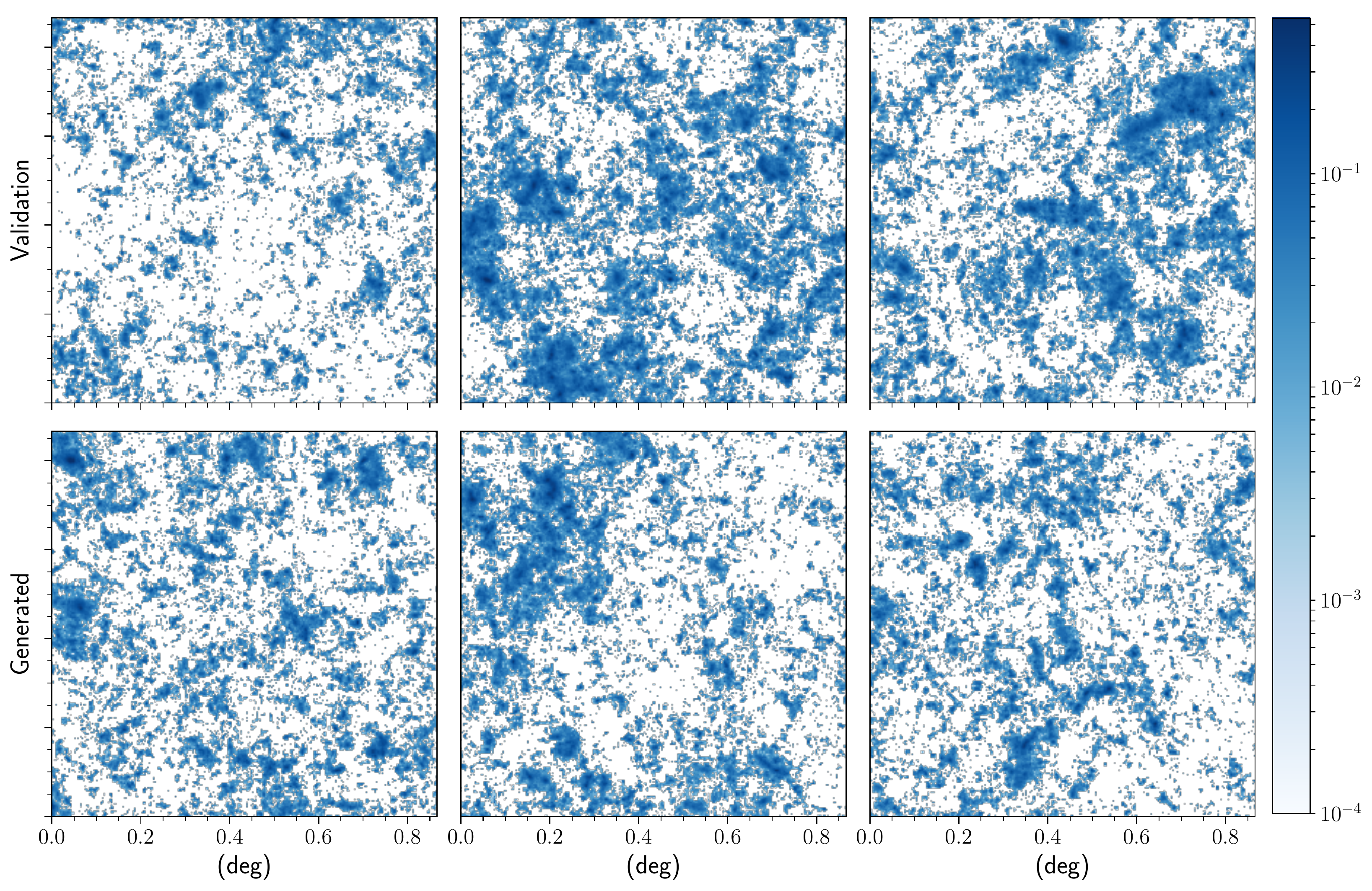}
    \caption{Examples of weak lensing convergence maps generated from
    outputs of cosmological $N$-body simulations (top) and those generated by CosmoGAN (bottom). Each panel shows the projected density distribution in about a square degree field. Dark parts are high density regions. 
    The density distributions (maps) in the top panels are generated from ray-tracing simulations of gravitational lensing
    using an array of outputs from direct
    $N$-body simulations. We show here three realizations of such simulations. CosmoGAN learns the
    features of the cosmological density distribution from a number of the simulated maps, and becomes able to generate images (bottom panels) 
    that resemble the "true" density distribution.
    }
    \label{fig:CosmoGAN}
\end{figure}

CosmoGAN has been developed to generate weak lensing convergence maps \cite{Mustafa19}. It is based on generative adversarial networks (GANs)
that are trained with outputs from a set of cosmological simulations. 
The generated images (lensing convergence fields) resemble those
made directly from $N$-body simulations
(see Figure \ref{fig:CosmoGAN}), 
and also reproduce an 
array of statistics from 1-point distribution function to Minkowski functionals with reasonable accuracy. 
Properly trained generative models have a potential to be replaced with costly direct simulations for well-defined purposes, but
it is worth noting that GAN models need to be trained with a vast
number of "true" samples, either from direct numerical simulations or
from observation. This has been possible so far for one or a few,
limited cosmological models, but the overall computational cost 
is still extremely large if one aims at developing a "universal" GAN 
that is capable of generating cosmological density fields for a broad range of theoretical models.

Standard techniques of density field reconstruction with WL yield noise dominated results. There are 
different kinds of noise associated with
real observations, and one of the important and intrinsic noise source
is the number density distribution of background galaxies. 
WL signals are measured
by detecting small coherent image distortion
of over tens to hundreds of galaxies within a
small area of the sky. By definition, the measurement error scales with the number of galaxies $N_{\rm gal}$ as $\propto 1/\sqrt{N_{\rm gal}}$. 
With the largest ground-based telescopes, the effective number of galaxies used for weak lensing measurement is a few tens in a 
1-${\rm arcmin}^2$
area \cite{Chang13}, and the finite number of galaxies is still a major source of noise for WL mass reconstruction. 

Several promising approaches have been proposed to denoise 
WL convergence maps using CNNs.
Ref.~\cite{Shirasaki21} use conditional GANs \cite{Isola16} 
to reconstruct convergence fields from noisy observational data.
The authors apply the GANs to real Subaru HSC data and show that the detected peak positions in the reconstructed map are matched well to the observed galaxy clusters identified by means of other observations.

\subsection{Galaxy redshift surveys}

Galaxy redshift survey \cite{Lahav04} is a straightforward way to map the three-dimensional distribution of galaxies. 
Unfortunately it requires time-consuming spectroscopic measurements of the target galaxies to determine the radial distance from the observer, and thus galaxy redshift surveys are more expensive than photometric imaging surveys. 
However, a simple statistics argument suggests that the information content extracted from a galaxy redshift survey can easily exceed that from an imaging survey. In practice, the shot noise contribution arising from the discrete nature of galaxy distribution restricts the full access to information on small scales. There is also a competition between the survey area and the depth, and the survey parameters are usually optimized to maximize the total signal-to-noise ratio. This optimization is done by considering multiple scientific goals in addition to the cosmological constraining power.

Theoretically, it is still difficult to predict accurately the distribution of galaxies using analytical or numerical methods. The relation between the underlying invisible dark matter and galaxies is called galaxy bias \cite{Desjacques18}.  Observationally, it is known that galaxies are biased with respective to dark matter differently depending on their luminosity, color or morphology. Therefore, accurate modeling of the distribution of a galaxy population can be done, if ever possible, only after the full criteria of the galaxy selection are specified. 
A common practice is to model or to parameterize the relation between dark matter and galaxies (see Sec. \ref{sec:galaxy_model} for ML-based methods), and then to compare the predicted summary statistics with those derived from observations.
A fast method to generate theoretical "templates" of the summary statistics using ML shall be discussed extensively in Sec.~5.
Such model templates can be tested by using, for instance, realistic mock galaxy catalogues generated from hydrodynamical simulations. 

For cosmology study, it is of crucial importance to select a population of galaxies whose properties are well understood. Luminous red galaxies (LRGs) or similar types of galaxies have been popular main targets of galaxy surveys in the past two decades \cite{Eisenstein01}. They are associated with massive halos, and the majority of them are located at the center of the host halo. Ongoing and forthcoming surveys such as DESI \cite{DESI16}, Subaru PFS \cite{PFS14}, and Euclid \cite{Euclid13} target emission line galaxies to map 
the galaxy and matter distribution at high redshifts. Clearly, understanding the nature of the emission line galaxies is important in order to fully utilize the data and to extract cosmological information from them in an unbiased manner. Also, novel ML-based methods would be needed to perform cosmological parameter inference beyond the traditional analyses based on 
low-order statistics.
ML approaches using three-dimensional CNNs \cite{Ntampaka20}
or graph neural networks (GNNs) have already been proposed. For the latter,
two nearby galaxies (nodes) are connected by edges \cite{Villanueva-Domingo22b}. 

\subsection{Intensity mapping}

The cosmic microwave background (CMB) radiation contains rich information on the evolution of the universe from the very early epoch just after the Big Bang to the present.
Past CMB experiments achieved accurate determination of  cosmological parameters \cite{Planck18}, and
future experiments including the Simons Observatory \cite{SimonsObs}, 
the CMB-S4 \cite{CMB-S4}, and LiteBIRD \cite{Sugai20} will perform polarization measurements
and multi-band intensity mapping with higher sensitivities. 

Intensity mapping measures everything altogether from near to far in a single band or in a wavelength pixel, and both extragalactic and Galactic foregrounds are unavoidably detected.
Refs. \cite{Troster19,Han21} propose to use generative models  
to predict the foreground components including
the kinetic and thermal Sunyaev-Zel’dovich effects, cosmic infrared background, and radio galaxies.
The authors show that the intensity power spectra and other non-linear statistics are reproduced from their mock observation.
Many realizations of wide-area, high-resolution maps generated by ML methods can be used to study potential systematic errors and to evaluate covariance matrices, which are crucial for precise cosmological analysis.
The use of conditional generative models are also proposed for other purposes such as removing the foreground \cite{Wang22_CMB}, 
separation of each component \cite{Bonjean22}, reconstruction of lensing map \cite{Caldeira19} and in-painting of masked regions \cite{Yi20,Puglisi20,VafaeiSadr21,Montefalcone21}.
To utilize all-sky data, one can extend a traditional 2D CNN to be applied to images on a sphere. 
Ref. \cite{Petroff20}  demonstrate that such a spherical neural network can be used to directly remove noises from all-sky CMB data.

\subsection{Line intensity mapping}

Line intensity mapping (LIM) is a rapidly developing observational technique to probe the large-scale structure by measuring collective fluctuations 
of line emissions in a broad range of wavelength.
LIM probes three-dimensional distribution of line emitters over a larger volume with smaller observational cost than galaxy redshift surveys. 
 Ongoing experiments are aimed at measuring the fluctuations of the 21-cm spin-flip transition line from neutral hydrogen \cite{Chang10},
and several large programs such as SKA \cite{Carilli15} and CHIME \cite{Amiri17} are also underway.
LIM observations targeting at different line emission from CO, H$\rm \alpha$, and Lyman-$\rm \alpha$,  
are also being conducted and planned \cite{Kovetz19}.
NASA's SPHEREx satellite to be launched in 2024 is expected to map infrared intensities over the full sky \cite{Dore18}. 

Refs. \cite{Zamudio-Fernandez19,Wadekar21} trained GANs using the outputs of IllustrisTNG hydrodynamics simulation in order to generate neutral hydrogen distributions at  post-reionization epochs ($z$ < 6). 
The resulting power spectrum and bispectrum from their models, HIGAN and HInet, agree with those of the original simulation outputs better than an analytical method based on halo occupation distributions.
Ref. \cite{Hassan21} proposes another type of generative model, HIFlow, 
which generates neutral hydrogen maps for a given set of input cosmological parameters. The authors train HIFlow using the simulation data from CAMELS (see Sec.~\ref{sec:galaxy}), to promote learning the diversity of intensity maps. Thus generated IM catalogues can be used for statistical analysis and for cosmological parameter inference with 21-cm LIM observations. 

Observations of the 21-cm line emission from the epoch of reionization (EoR) can be a direct probe of cosmic reionization.
Because the spatial and frequency distributions of the 21-cm emission are highly non-linear
and thus difficult to interpret, 
ML could be the most effective way to extract cosmological information.
Various ML methods have been proposed to infer cosmological and astrophysical parameters 
from the intensity power spectrum \cite{Shimabukuro17,Doussot19,Choudhury22}, 
other summary statistics \cite{Jennings20,Choudhury21},
and directly from tomographic maps \cite{Gillet19,Hassan19,Neutsch22} of the 21-cm line intensity.
Ref.~\cite{Hassan19} showed that a ML model can distinguish reionization models even when 
a realistic noise level of SKA is assumed.
Other ML applications to 21-cm observations include 
identification of ionized regions \cite{Bianco21},
reconstruction of 21-cm maps from Lyman-$\rm \alpha$ emitter distributions \cite{Yoshiura21},
and reconstruction of DM distributions from 21-cm maps \cite{Villanueva-Domingo21}.
Emulation (see Section \ref{sec:emulation}) and generation \cite{List20} of the EoR 21-cm signals have also
 been explored.

\begin{figure}
    \centering
    \includegraphics[width=14cm]{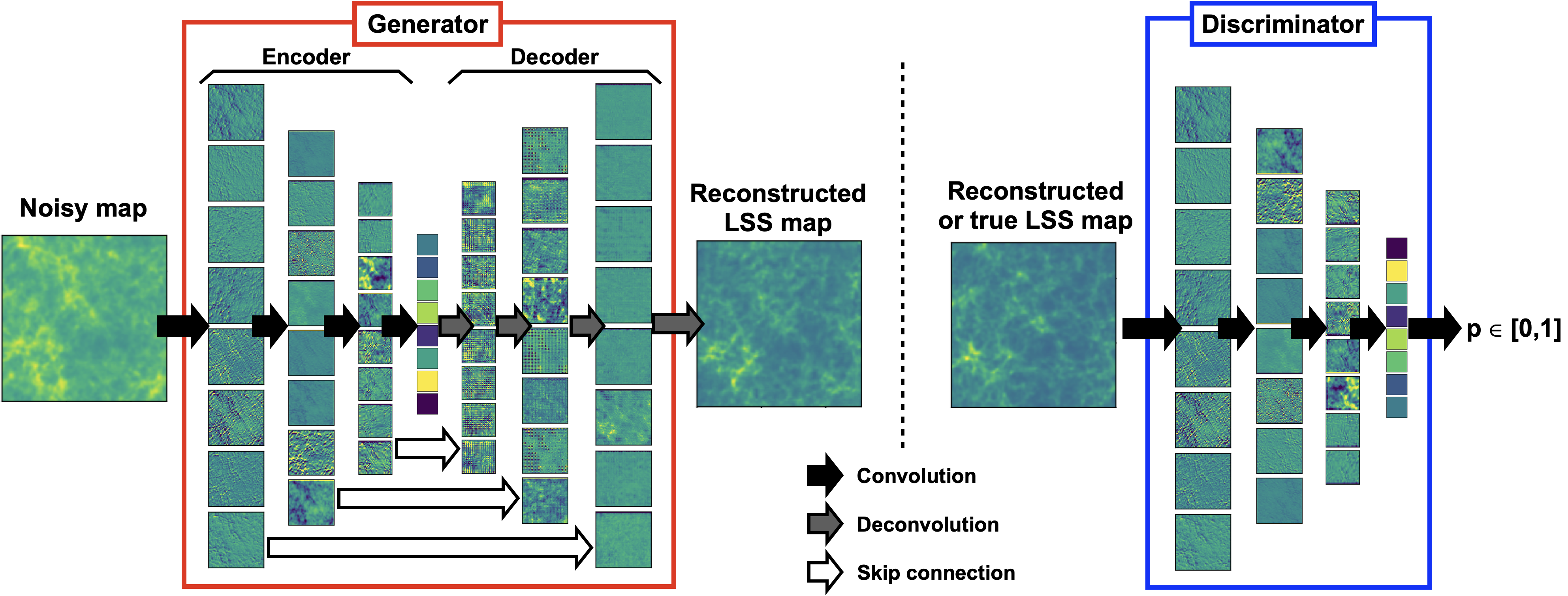}
    \caption{Schematic picture of a conditional GAN used in Ref. \cite{Moriwaki20}  
    with two CNNs called generator (left) and discriminator (right). 
    The generator reconstructs a LSS map from a noisy observational map
    while the discriminator is given either reconstructed or true map and returns a probability $p\in [0,1]$ that the input is true map.
    The CNNs consist of several convolutional and deconvolutional layers (black and gray arrows), where the input images are convolved with various filters. 
    The generator has a so-called skip connection (white arrow) \cite{Isola16}, where the outputs in the encoder layers are reused in the decoder layers.
    The training is done adversarially; the discriminator is updated to distinguish between reconstructed (generated) and true data more accurately,
    while the generator is updated to better {\it fool} the discriminator.
    Eventually the generator becomes able to reconstruct very complex features.
    }
    \label{fig:cGAN}
\end{figure} 

Foreground and background contamination is a serious problem of LIM. 
Conditional generative models can be used for information extraction and component separation from LIM data.
In 21-cm observations, emissions from our Galaxy and extragalactic radio sources are
much brighter than the high-redshift 21-cm signals.
Ref. \cite{Li19} proposes to remove foreground emissions from 21-cm observation data.
The authors utilize an autoencoder, a popular generative CNN model,
to reconstruct signals from the EoR from noisy observational data assuming SKA.
In LIM observations in other wavelengths, there are multiple bright emission lines,
and the so-called interloper lines compromise reconsruction of the LSS to be traced by the target line emission.
Removal of line interlopers can done by trained CNNs
\cite{Moriwaki20}.
The authors generate mock observational data assuming SPHEREx mission
and use conditional GANs (Figure \ref{fig:cGAN}) to separate two emission lines H$\rm \alpha$ and [O{\sc iii}]. 
Three dimensional data (angular $\times$ spectral domain) that are obtained by LIM observations 
can also be analyzed with three-dimensional CNNs \cite{Makinen21,Moriwaki21} or with recurrent neural networks \cite{Prelogovic22}.
Including the additional dimension is expected to improve the accuracy of reconstruction and regression.

\section{From simulation to emulation}
\label{sec:emulation}

Numerical simulations have played
a vital role in establishing the basic picture of how the observed large-scale structure is formed (see \cite{Angulo22} for a recent review). 
The so-called standard cosmological model 
allows us to
generate accurate initial conditions that reproduce the statistical
properties of the early Universe.
Hence, the remaining task 
in our {\it forward modeling} is to implement
appropriate physical processes and to follow the structure formation and 
evolution within a sufficiently
large volume that represents at least a good fraction of the observable universe.
This fact grants cosmological simulations
to be particularly useful for making theoretical predictions.

Theoretical models are tested by performing detailed comparisons with a broad range of observations. Bayesian statistics is the fundamental framework to derive the region of model parameters and/or select a model that are consistent with observational data. However, due to the high computational cost, direct application of simulations to statistical inference problems is not straightforward.
Obviously one cannot generate a sufficiently large number
of realizations by performing structure formation simulations to incorporate in the standard statistical procedures such as Markov chain Monte Carlo (MCMC). 

Often, summary statistics are used instead of the observed matter/galaxy distributions, which are usually composed of multivariate random variables. We can significantly reduce the volume of the data vector without much loss of information by identifying appropriate summary statistics, making a fully statistical inference feasible. In this context, numerical simulations can be used to calibrate the response of the statistics to the model parameters. One popular methodology for performing this task is emulation. Costly numerical simulations are replaced with a cheaper but accurate statistical model, which can then be used directly in the inference process. In this section, we show how this idea has been implemented and employed in recent cosmological inference problems.

\subsection{Cosmological parameter estimation}
Cosmological parameters are accurately inferred from 
a variety of observations, including weak lensing \cite{Joudaki17,Hikage19,Asgari21,Secco22} and galaxy clustering \cite{2dFGRS05,WiggleZfinal,6dFGRSRSD,FMOSRSD,VipersRSD,SDSSDR12,SDSSDR16}. 
It is becoming increasingly more popular to incorporate
ML into some elements in the cosmological parameter inference procedure (e.g., \cite{Gupta18,Ribli19a,Ravanbakhsh16,Mathuriya18,Gillet19,Hassan19}).

In modern cosmology, the basic parameters such as the
expansion rate of the present-day Universe and the matter contents are inferred based on Bayes' theorem
\begin{equation}
    P ({\bm \theta} | {\bm d}) \propto  P ({\bm d} | {\bm \theta}) P ({\bm \theta}) 
\end{equation}
where the left-hand side gives the probability distribution
of the model parameter(s) ${\bm \theta}$ conditioned on
data ${\bm d}$ (the posterior). To perform the inference, one needs to know
$P ({\bm d} | {\bm \theta})$, the distribution of ${\bm d}$ for a model specified by ${\bm \theta}$ called likelihood, and the prior model distribution $P ({\bm \theta})$.
In many physics applications, forward modeling is possible using either analytic or perturbation calculations or by
direct numerical simulations. Unfortunately, evaluating
the likelihood $P ({\bm d} | {\bm \theta})$ is often 
extremely time consuming and impractical, because the
model variation is usually large; the parameter space ${\bm \theta}$ is multidimensional and the typical curse of dimension
kills essentially all brute-force attempts.

A partial remedy for the difficulty in large-scale structure cosmology is that an unbiased estimator for reasonable summary statistics, such as the two-point correlation function or its Fourier counterpart, the power spectrum, are known and understood well. Also the respective estimator follows a normal distribution with good accuracy in most of the cases (but see, e.g., \cite{Sato11,Simon15,Sellentin18} for potential violation of Gaussianity in case of weak lensing statistics). This nice feature follows from the central limit theorem, together with the fact that such estimators are often written by a combination of many random variables. What we need in practice are the expectation values of the statistics and the covariance matrices. Unfortunately, estimating the covariance matrix is still a challenging task to do in a direct manner using numerical simulations, even for a single cosmological model \cite{Takahashi09,Blot15}. It typically requires {\it at least} a few hundreds realizations of either fully nonlinear simulations or those with some approximations. Even with the advent of modern supercomputers, 
it is challenging to perform all necessary direct simulations at all the parameter sampling points. 
Interesting discussions are found on the consequences of the cosmology-independent covariance approximation for different observational probes \cite{Eifler09,Labatie12,Carron13,Kodwani19,HarnoisDeraps19}.

In this section, we introduce an efficient emulation-based approach to circumvent this problem based on a Gaussian likelihood with a cosmology-independent covariance matrix. 
We would like to mention here briefly that there has been an impressive progress in performing
parameter inference based on the comparison between numerical simulations and observations (see \cite{Cranmer20} for a review), or based on new formulations of the likelihood function in order to deal with the observed random fields without bypassing summary statistics. 
The former includes methods designed to avoid the computationally expensive evaluation of likelihood, commonly known as likelihood-free inference.
One notable example is Approximate Bayesian Computation \cite{Rubin84}, which has already been applied in
several cosmological studies \cite{Weyant13,Ishida15,Akeret15,Lin16,Hahn17}.
The latter approach is often called as field-level inference \cite{2020JCAP...01..029E,2022JCAP...08..003T,2022MNRAS.509.3194P,2022arXiv220413216S,2023MNRAS.520.5746A,2022JCAP...08..007B,2021JCAP...04..032S,2021JCAP...01..067C,2020JCAP...11..008S,2020JCAP...04..042C,2019A&A...621A..69R},
which has the potential to maximally extract the information contained in the observed universe, by exploring a huge parameter space that includes not only cosmological parameters but also a set of Fourier modes of the random initial condition, often aided by an efficient Hamiltonian Monte Carlo sampler (\cite{HMC} and see \cite{2010MNRAS.407...29J,2010MNRAS.409..355J,2013MNRAS.432..894J} for early attempts in cosmological large-scale structure), differential programming (\cite{10.1007/978-3-319-55696-3_3,JMLR:v18:17-468,10.5555/3327546,Innes2018,2019arXiv190707587I} and \cite{Modi21,2021A&C....3600490B,2022arXiv221109815L} for cosmological simulations), or optimization in place of sampling \cite{2017JCAP...12..009S,2018JCAP...07..043F,2021JCAP...10..056M,2021arXiv210412864M}. For both the methods, developing a fast simulator is a crucial task. ML techniques are also used for this purpose, by partly replacing the gravitational force calculation in $N$-body simulations \cite{2022arXiv220705509L} or by improving the accuracy by correcting for the mistakes in approximate dynamics or by super-resolution emulation \cite{2019PNAS..11613825H,2020MNRAS.495.4227K,2021MNRAS.507.1021N,2021PNAS..11822038L,2022ApJ...930..115K}. 

The emulator approach based on summary statistics discussed in what follows is relatively conservative in the sense that it is fully based on the traditional Bayesian inference pipeline with ML elements introduced only in the prediction of the expectation values of the summaries. Nevertheless it has already been shown that such approaches achieve improvements in constraining cosmological parameters compared to traditional methods. On top of this, the advances in the inference methodology, in particular the application of ML techniques, provide a promising avenue for a more efficient cosmology analysis.

\subsection{Emulation}

A practical way to bypass time-consuming numerical simulations in Bayesian inference problems is to replace it with an emulator. Here, we introduce emulators which predict summary statistics given a set of model parameters, instead of generating fully three-dimensional cosmological density fields. The outputs from an emulator serve as accurate templates in parameter inference problems.

Let us take the density fluctuation power spectrum as a frequently used summary
statistics in cosmology. 
Conventionally, functional fits were often used
\cite{Hamilton91,Peacock94,Smith03}.
Fitting formulae have been proposed using analytic expressions of a set of functions. Although such attempts are often motivated by physical considerations, they employ empirical analytic forms that are validated and with the internal parameters calibrated against numerical simulations. However, there is a tough demand for the accuracy to match those of modern and future cosmology surveys \cite{DETF,2022JHEAp..34...49A}. It is difficult to keep up with and update the model constantly, even if a larger set of simulation data is becoming available. Although recalibration of the model parameters can be performed fairly straightforwardly \cite{Takahashi12}, the complexity of the analytical form to be used eventually limits the best achievable accuracy. In addition, generalization of an existing fitting formula to allow more input cosmological parameters is a non-trivial task \cite{Mead15}.

The concept of an emulator is fundamentally different from 
that of conventional methods that utilize interpolation 
and/or extrapolation of summary statistics. An emulator is a statistical model that deals with a regression problem. It learns the relation between inputs and outputs obtained by simulations. The task of an emulator is to generalize the relation obtained at a finite number of sampling points to a new input selected at an arbitrary point in the input parameter space of interest. Modern emulators are based on Bayesian inference of physical quantities. It is often constructed in a non-parametric manner, with the complexity automatically calibrated in a data-driven way. This is the most remarkable feature in contrast to traditional functional fitting, for which one has to manually determine the functional form. 

The advantage of constructing emulators over other simulation-based inference models is that they are {\it amortizable} in the following sense. First, one can construct an emulator without actual observational data, with which statistical inference would eventually be performed. Therefore, the emulator construction task can be started \textit{before} an observational program is started, or before its details such as the depth, area, or full specification of the target galaxies are fixed. Once an emulator is developed, it can be used for different purposes, not only for statistical inference from the observational program originally in mind,
but also for understanding, for instance, which input parameter (or which combination of different parameters) is important. The result can be used, in turn, to optimize future observations. Unlike the emulator approach, direct applications of a simulation-based statistical model in inference is non-amortizable: the generated simulation data are only for the specific inference problem. The existence of unbiased estimators of standard statistical measures in large-scale structure cosmology makes emulation-based summary statistics templates useful and generic.

The idea of emulation was first introduced to the problem of cosmological structure formation by~\cite{Heitmann06,Habib07}. Over more than ten years since then, there are more than several research groups who advance the idea with different methods for different purposes. We will discuss the key ingredients for successful emulators and their applications to actual data from the recent literature.

\subsubsection{The design of experiments}
The first important step is efficient arrangement of simulations in the input parameter space. Often, one has to deal with a multiple regression problem in which the inputs have to be sampled in a high-dimensional parameter space. The design of experiments (DoE) itself is of common research interest in different domains of science and engineering (see \cite{DoE} for a review). This is a typical exploration-exploitation trade-off problem: uniform coverage over the possible domain in the input parameter space is desirable, on the one hand, while prioritizing the parameter regions where previous observations fit better could be more suited. In the case of emulator development, one usually places more weight on exploration and prepares a database which covers a wide parameter space such that the high posterior probability region from an unseen future observation would be included. Therefore, a uniform sampling scheme over the target parameter space is useful for this problem.

\begin{figure}
    \centering
    \includegraphics[width=13cm]{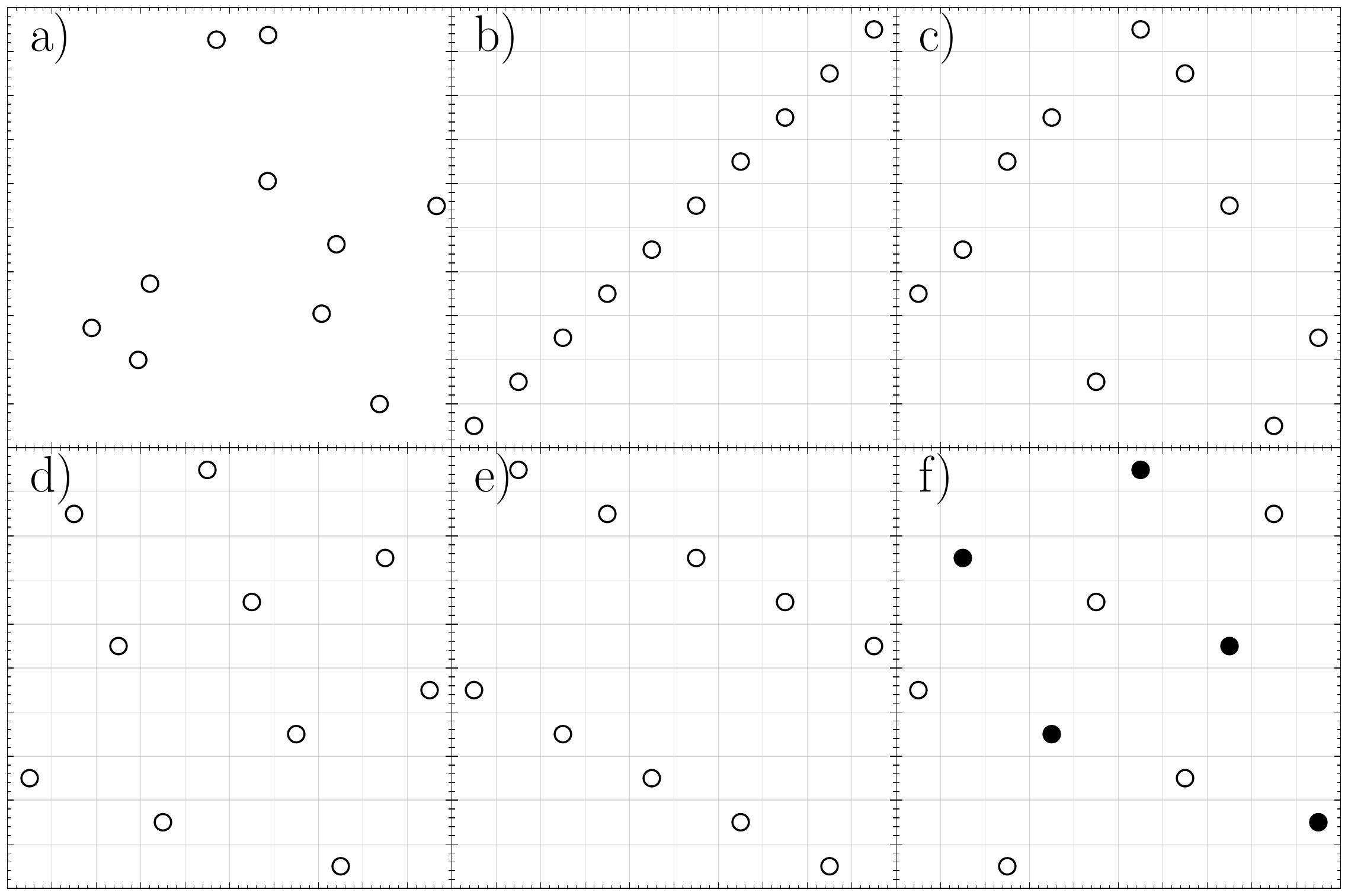}
    \caption{Comparison of DoEs with ten points in a two dimensional space. a) Random uniform distribution, b) -- f) Latin Hypercube Design (LHD), i.e., one and only one data points in every row and column. b) Diagonal, e) Random LHD, d) Maximin distance LHD, e) Maximin distance LHD, but with an anisotropic distance metric, f) Sliced maximin distance LHD \cite{Ba15} with two slices: Each slice depicted by different symbols. The whole data points, as well as those in the same slice, form an LHD.}
    \label{fig:LHD}
\end{figure}

There are different indicators of good DoE. The first is the space-filling property. One usually wants to avoid placing a new input parameter set close to a point where a simulation has already been performed. Also, one does not have a large gap around a certain location of the parameter space over which simulation data do not exist. The second is the projection property. One can imagine a situation where one of the input parameters has little impact on the outcome of the experiment. In this case, the lower-dimensional subspace after projection of this unimportant parameter should be filled with desirable properties.
Having these two properties together makes a DoE suitable for ``black-box'' problems, where the relation between input and output can only be known after performing simulations. A popular choice is the Latin-Hypercube Design (LHD; \cite{McKay79} and see some examples in Figure~\ref{fig:LHD}). The good space-filling property is ensured by applying conditions such as orthogonality or the maximin distance condition\footnote{To maximize the minimum distance between all the pairs of design points. This ensures that the samples tend to become more distant and fill the whole space of interest evenly.}.
Another aspect of a DoE is granularity. A one-time DoE with good space-filling and projection properties could be difficult to expand by increasing the number of sampling points while maintaining the original good property. LHD is usually not great in this regard, as there is no natural way to expand.

\subsubsection{Regression models}

Gaussian Process Regression (GPR) builds a non-parametric model
based on available data. The valuable features are that a GPR returns 
the associated errors together with the central values \cite{Rasmussen06}.

The central assumption of GPR is that the output of the unknown function, $f(\boldsymbol{x})$, follows a normal distribution. One can imagine that $\boldsymbol{x}$ stands for cosmological parameters and $f$ is the matter power spectrum at some wavenumber $k$. We introduce the probability density functional of this function, $P[f(\boldsymbol{x})]$. The Gaussianity assumption implies that this functional is fully characterized by two functions, the mean function $\langle{f(\boldsymbol{x})}\rangle$ and the covariance function $C(\boldsymbol{x}_1,\boldsymbol{x}_2) = \left\langle{(f(\boldsymbol{x}_1)-\langle{f(\boldsymbol{x}_1)}\rangle)(f(\boldsymbol{x}_2)-\langle{f(\boldsymbol{x}_2)}\rangle)}\right\rangle$, where the operation $\langle\dots\rangle$ denotes the average. The covariance function is also commonly called the kernel function. 

A key point here is that, while the relation between the cosmological parameters and the power spectrum is deterministic in principle, one introduces a statistical argument that says, due to the lack of full knowledge of the function, or sufficiently dense samples of simulation data, we are not certain about the value of the power spectrum at given cosmological parameters. We can still introduce our prior knowledge by assuming certain functions for the mean and covariance. The common practice is that the mean function is set to zero, unless one has a good idea of the expected behavior of the function $f$. Then, the covariance function sets the typical amplitude and the flexibility of the matter power spectrum as a function of cosmological parameters. The idea of GPR is to shrink the uncertainty intervals by adding simulation data. This can be done naturally considering the conditional probability $P[f(\boldsymbol{x}_{N+1})|\boldsymbol{Y}_N]$ in the presence of data $\boldsymbol{Y}_N = \{y_i|i=1,\cdots,N\}$ obtained by simulations conducted at sampling points $\boldsymbol{X}_N = \{\boldsymbol{x}_i|i=1,\cdots,N\}$. One can choose $\boldsymbol{x}_{N+1}$ at the new input point at which one would like to make a prediction in the presence of $N$ data points. Thanks to Gaussianity, the mean and variance of this conditional probability can be computed analytically: 
\begin{eqnarray}
\langle y^{(N+1)}|\boldsymbol{Y}_N\rangle =  \boldsymbol{k}^T\boldsymbol{C}_N^{-1}\boldsymbol{Y}_N,
\label{eq:GP_mean}
\end{eqnarray}
and
\begin{eqnarray}
\langle (\Delta y^{(N+1)})^2|\boldsymbol{Y}_N\rangle =
\kappa - \boldsymbol{k}^T\boldsymbol{C}_N^{-1}\boldsymbol{k},
\label{eq:GP_var}
\end{eqnarray}
respectively, where we have introduced an $N$-by-$N$ matrix $\boldsymbol{C}_{N} = \{C(\boldsymbol{x}_i,\boldsymbol{x}_j) | 1 \leq i, j \leq N\}$, which gives the covariance between the existing input points, an $N$-element vector $\boldsymbol{k} = \{C(\boldsymbol{x}_i,\boldsymbol{x}_{N+1}) | 1 \leq i \leq N\}$ between the existing and the new input, and the variance at the new input $\kappa = C(\boldsymbol{x}_{N+1},\boldsymbol{x}_{N+1})$.
It is easy to see in Eqs.~(\ref{eq:GP_mean}) and (\ref{eq:GP_var}) that without any simulation data, the mean is zero and the variance is $\kappa$ by construction. The presence of simulation data, $\boldsymbol{Y}_N$, shifts the mean and shrinks the variance according to these equations. All of these computations are written as matrix products and thus are easily evaluated. A possible bottleneck of this model is the inversion of the matrix for $\boldsymbol{C}_N^{-1}$, which can be expensive when the size of the data $N$ increases.

\begin{figure}
    \centering
    \includegraphics[width=13cm]{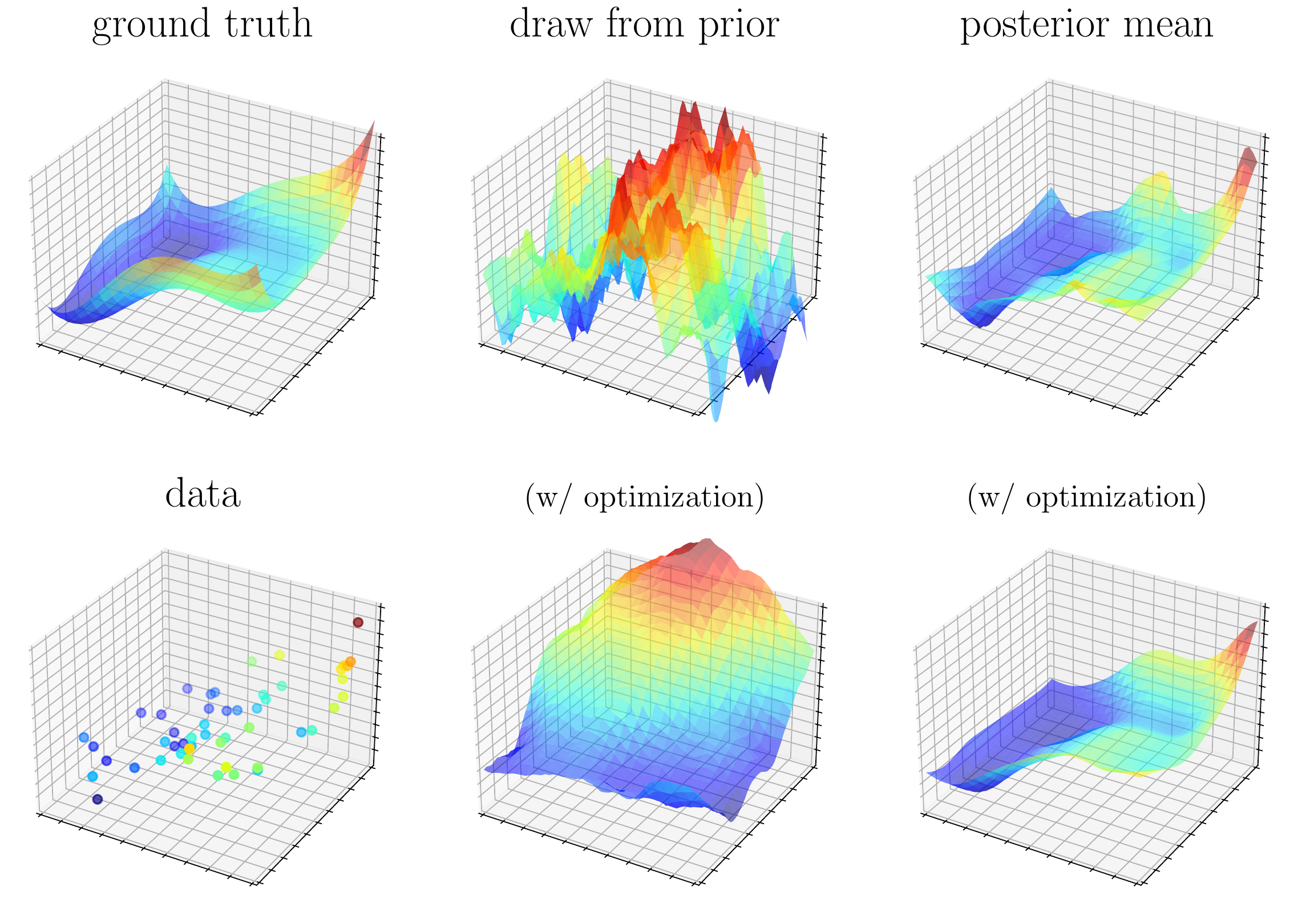}
    \caption{Example application of GP to a two inputs-one output regression problem. The task is to infer the 2D surface in the upper left panel from the 40 data points with random noise in the lower left panel. The two middle panels depict random draws from GP assuming the Mat\'ern 32 kernel function, before (upper) or after (lower) the parameter optimization on the data. One can see that the smoothness and the typical level of fluctuations of the surface is closer to the ground truth after the optimization. The right panels show the expectation value of the Gaussian probability density corresponding to the middle panels, now conditioned on the data points in the lower left panel. A GP with a fixed kernel function works as the prior and the final outcome can be regarded as the posterior in the standard Bayesian method.}
    \label{fig:GP}
\end{figure}

The remaining task for us is to specify the covariance function $C(\boldsymbol{x}_1,\boldsymbol{x}_2)$. One can design this function according to the prior knowledge of the target function $f(\boldsymbol{x})$: If one expects that this function should exhibit periodicity, for instance, one may try a periodic function for $C(\boldsymbol{x}_1,\boldsymbol{x}_2)$. A common practice, without any prior knowledge on $f(\boldsymbol{x})$ is to consider a \textit{stationary} kernel, that is, $C(\boldsymbol{x}_1,\boldsymbol{x}_2)=C(|\boldsymbol{x}_2-\boldsymbol{x}_1|)$, solely determined by the distance between the two inputs. This greatly reduces the number of free parameters that characterize the covariance function. Often, a simple function is employed, such as a Gaussian-form radial basis function, an exponential, or the Mat\'ern kernel. The free parameters for these kernels determine the amplitude and the distance metric in the multidimensional input space. A unique aspect of GP is that one can ``learn'' this function from the data $\boldsymbol{Y}_N$, by introducing free parameters in $C(\boldsymbol{x}_1,\boldsymbol{x}_2)$ and maximize the probability of having $\boldsymbol{Y}_N$. In this sense, this rather traditional Bayesian framework can be regarded as ML. Each step of GP application to a simple example problem can be found in Figure~\ref{fig:GP}.

The possible drawback of a GP is that multiple output regression is not straightforward: We have focused on the simplest case of single output functions, $f(\boldsymbol{x})$. Extensions for correlated multi-output problems are usually not straightforward. In the typical applications to cosmology, such as the power spectrum emulation, it is natural to consider a multiple-output model and predict the function values at various wavenumbers when a set of cosmological parameters are specified as the input. Since the size of the full covariance function between different input-output combinations can be prohibitively large, one has to find an efficient way to decompose it or connect the final output to intermediate variables generated by a smaller model in the latent space \footnote{Multiple output Gaussian process is implemented for instance in a public PYTHON package, \textsc{Gpy} (https://sheffieldml.github.io/GPy/).}. To avoid the complexity of multiple output regression, a popular workaround is to reduce the dimensionality of the outputs by methods such as singular value decomposition or principle component analysis, to convert to a smaller number of less-correlated data vector. After this operation, each of the vector element is modeled by an independent GP, ignoring the cross correlation. Alternatives to GP in this context include fully-connected feedforward neural networks (FFNN) and the Polynomial Chaos Expansion \cite{10.2307/j.ctv7h0skv}\footnote{Polynomial Chaos Expansion (PCE) is an efficient approximator of a target (multivariate) function by using a finite set of polynomials which form an orthogonal basis with respect to the distribution of the input variables.}. Since the dependence of summary statistics on the cosmological parameters is typically smooth and only weakly nonlinear, unlike the underlying stochastic field realizations, an FFNN with rather smaller number of hidden layers or low-order polynomial expansion are known to work accurately. The optimal regression model should depend on the precise definition of the problem. Nevertheless, a thorough comparison among different models is still missing and would be useful to prepare for the next generation observational programs (see, e.g.,~\cite{Schneider11,Angulo21,Cuesta-Lazaro22} for studies along this line).

\subsection{Cosmological emulators and application to real data analysis}

\begin{table}
\caption{Structure formation simulation campaigns aimed at cosmological emulators. We summarize the cosmological parameter space (flatness assumed in all the cases), the design of experiments (acronyms: LH = Latin Hypercube, Mm = Maximin distance), the statistical model used for regression (FIT = function fitting, PC = Principal components; we do not distinguish Singular Value Decomposition from PC analysis here, GP = Gaussian Process, NN = Neural Network, PCE = Polynomial Chaos Expansion, LIN = first-order Taylor expansion around the fiducial model), the target statistics of regression, the emulated statistics, and the relevant papers, including both descriptions of simulations and regression modeling.\label{tab:emulator}}
\centering
\resizebox{\textwidth}{!}{
\begin{threeparttable}
 \begin{tabular}{c ||c c c c c} 
  & Cosmology & \multirow{2}{*}{DoE} & Regression & \multirow{2}{*}{Statistics} & \multirow{2}{*}{Reference}\\ 
  & (parameters) &  & model &  & \\ 
 \hline\hline
 \multirow{2}{*}{Coyote Universe} & $w$CDM & \multirow{2}{*}{symmetric LH} & \multirow{2}{*}{PC|GP} & $P_\mathrm{m}(k,z|\boldsymbol{\theta}_\mathrm{cosmo})$ & \cite{Heitmann10,Heitmann09,Lawrence10,Heitmann14}\\ 
  & $(\omega_\mathrm{m}, \omega_\mathrm{b}, n_\mathrm{s}, w, \sigma_8)$ & &  & $c(M|\boldsymbol{\theta}_\mathrm{cosmo})$ & \cite{Kwan13}\\
 \hline
 \multirow{2}{*}{PkANN} & $\nu w$CDM & \multirow{2}{*}{LH\tnote{a}} & \multirow{2}{*}{NN} & \multirow{2}{*}{$P_\mathrm{m}(k | \boldsymbol{\theta}_\mathrm{cosmo}, z)$} & \multirow{2}{*}{\cite{Agarwal12,Agarwal14}}\\
 & $(\omega_\mathrm{m}, \omega_\mathrm{b},n_\mathrm{s},w_0,\sigma_8,\sum m_\nu)$ & & & & \\
 \hline
 \multirow{2}{*}{Mira-Titan Universe} & $\nu w_0w_a$CDM & \multirow{2}{*}{nested lattice}  & \multirow{2}{*}{PC|GP} & $P_\mathrm{m}(k,z|\boldsymbol{\theta}_\mathrm{cosmo})$ & \cite{Heitmann16,Lawrence17,Moran22}\\
 & $(\omega_\mathrm{m}, \omega_\mathrm{b}, \sigma_8, h, n_\mathrm{s}, w_0, w_a, \omega_\nu)$ & & & $n_\mathrm{h}(M|\boldsymbol{\theta}_\mathrm{cosmo})$ & \cite{Bocquet20}\\
 \hline
 \multirow{6}{*}{MassiveNuS} & \multirow{5}{*}{$\nu\Lambda$CDM} & \multirow{6}{*}{LH\tnote{b}} & -- & -- & \cite{Liu18}\\
& \multirow{5}{*}{$(\sum m_\nu, \Omega_\mathrm{m}, A_\mathrm{s})$} &  & \multirow{5}{*}{GP} & $C_\ell^{(\kappa)}(\boldsymbol{\theta}_\mathrm{cosmo})$ & \cite{Liu19,ZackLi19} \\
&  &  &  & PDF$^{(\kappa)}(\kappa|\boldsymbol{\theta}_\mathrm{cosmo})$ & \cite{Liu19} \\
&  &  & & $b^{(\kappa)}_{\ell_1,\ell_2,\ell_3}(\boldsymbol{\theta}_\mathrm{cosmo})$ & \cite{Coulton19} \\
&  &  & & $N_\mathrm{peak}^{(\kappa)}(\mathrm{S/N}|\boldsymbol{\theta}_\mathrm{cosmo})$ & \cite{ZackLi19}\\
&  &  & & $V_i^{(\kappa)}(\nu|\boldsymbol{\theta}_\mathrm{cosmo})$ & \cite{Marques19}\\
 \hline
 \multirow{4}{*}{cosmo-SLICS} & \multirow{3}{*}{$w$CDM} & \multirow{4}{*}{MmLH} & \multirow{4}{*}{PC|GP} & $\xi_{\pm}(\theta|\boldsymbol{\theta}_\mathrm{cosmo})$ & \cite{HarnoisDeraps19}\\
 & \multirow{3}{*}{$(\Omega_\mathrm{m}, \sigma_8, h, w_0)$} & & & $N_\mathrm{peak}^{(\kappa)}(\mathrm{S/N}|\boldsymbol{\theta}_\mathrm{cosmo})$ & \cite{HarnoisDeraps21,Davies22}\\
 &  & & & PDF$^{(\kappa)}(\mathrm{S/N}|\boldsymbol{\theta}_\mathrm{cosmo})$ & \cite{Martinet21}\\
 &  & & & $\xi_\mathrm{peak}^{(\kappa)}(\theta|\boldsymbol{\theta}_\mathrm{cosmo})$ & \cite{Davies22}\\ 
 \hline
 \multirow{4}{*}{EuclidEmulator} & $w$CDM & \multirow{4}{*}{MmLH} & \multirow{4}{*}{PC|PCE} & \multirow{4}{*}{$P_\mathrm{m}(k,z|\boldsymbol{\theta}_\mathrm{cosmo})$} & \multirow{2}{*}{\cite{EuclidEmu1}}\\ 
  & $(\omega_\mathrm{b}, \omega_\mathrm{m}, h, n_\mathrm{s}, w_0, \sigma_8)$ &  &  &  & \\ 
 & $\nu w_0w_a$CDM &  &  &  & \multirow{2}{*}{\cite{EuclidEmu2}}\\ 
  & $(\Omega_\mathrm{b}, \Omega_\mathrm{m}, \sum m_\nu, n_\mathrm{s}, h, w_0, w_a, A_\mathrm{s})$ &  &  &  & \\ 
  \hline
 \multirow{6}{*}{Aemulus} & \multirow{5}{*}{$w$CDM+$N_\mathrm{eff}$} & \multirow{6}{*}{LH\tnote{c,d}} & -- & -- & \cite{DeRose19}\\ 
  & \multirow{5}{*}{$(\omega_\mathrm{b},\omega_\mathrm{c},w_0,n_\mathrm{s},\ln(10^{10}A_\mathrm{s}),h,N_\mathrm{eff})$} &  & FIT|GP & $n_\mathrm{h}(M,z|\boldsymbol{\theta}_\mathrm{cosmo})$ & \cite{McClintock19}\\  
  &  &  & GP & $w_\mathrm{p}(r_\mathrm{p}|\boldsymbol{\theta}_\mathrm{cosmo},\boldsymbol{\theta}_\mathrm{galaxy})$ & \cite{Zhai19,Zhai22}\\  
  &  &  & GP & $\xi_{0,2}(s|\boldsymbol{\theta}_\mathrm{cosmo},\boldsymbol{\theta}_\mathrm{galaxy})$ & \cite{Zhai19,Zhai22}\\  
  &  &  & FIT|GP & $b(M,z|\boldsymbol{\theta}_\mathrm{cosmo})$ & \cite{McClintock19b}\\  
  &  &  & PC|PCE & $P_{ij}(k|\boldsymbol{\theta}_\mathrm{cosmo})$ & \cite{Kokron21}\\  
 \hline
\multirow{3}{*}{AbacusCosmos} & \multirow{2}{*}{$w$CDM} & \multirow{3}{*}{MmLH\tnote{d}} & -- & -- & \cite{Garrison18}\\
 & \multirow{2}{*}{$(\omega_\mathrm{b},\omega_\mathrm{m},h,n_\mathrm{s},\sigma_8,w_0)$} & & \multirow{2}{*}{GP} &  $\Delta\Sigma(r_\mathrm{p}|\boldsymbol{\theta}_\mathrm{galaxy},\boldsymbol{\theta}_\mathrm{galaxy})$ & \multirow{2}{*}{\cite{Wibking20}}\\
& & & & $w_\mathrm{p}(r_\mathrm{p}|\boldsymbol{\theta}_\mathrm{galaxy},\boldsymbol{\theta}_\mathrm{galaxy})$ & \\
\hline
 \multirow{3}{*}{AbacusSummit} & \multirow{2}{*}{$w_0w_a$CDM$+N_\mathrm{eff}$+running} & \multirow{2}{*}{ellipsoidal surface\tnote{e}} & -- & -- & \cite{Maksimova21}\\
& \multirow{2}{*}{$(\omega_\mathrm{b},\omega_\mathrm{c},n_\mathrm{s},\sigma_8,w_0,w_a,\alpha_\mathrm{s},N_\mathrm{eff})$} & \multirow{2}{*}{+ uniform random} & LIN & $P_{ij}(k|\boldsymbol{\theta}_\mathrm{cosmo})$ & \cite{Hadzhiyska21}\\ 
&  &  & GP & $\xi_\mathrm{g}(r_\mathrm{p},\pi|\boldsymbol{\theta}_\mathrm{cosmo},\boldsymbol{\theta}_\mathrm{galaxy})$ & \cite{Yuan22}\\ 
 \hline
 \multirow{2}{*}{BACCO} & $\nu w_0w_a$CDM & \multirow{2}{*}{LH\tnote{f}} & PC|GP or NN & $P_\mathrm{m}(k,z|\boldsymbol{\theta}_\mathrm{cosmo})$ & \cite{Angulo21}\\ 
  & $(\sigma_8,\Omega_\mathrm{m},\Omega_\mathrm{b},n_\mathrm{s},h,\sum m_\nu,w_0,w_a)$ & & NN & $P_{ij}(k|\boldsymbol{\theta}_\mathrm{cosmo})$ & \cite{Zennaro21}\\ 
 \hline
 \multirow{4}{*}{Dark Quest} & \multirow{3}{*}{$w$CDM} & \multirow{4}{*}{sliced MmLH} & \multirow{2}{*}{PC|GP or NN} & $\xi_\mathrm{hm}(x,z,M|\boldsymbol{\theta}_\mathrm{cosmo})$ & \multirow{3}{*}{\cite{Nishimichi19,Cuesta-Lazaro22}}\\ 
  & \multirow{3}{*}{$(\omega_\mathrm{b}, \omega_\mathrm{c}, \Omega_\mathrm{de}, \ln(10^{10}A_\mathrm{s}), n_\mathrm{s}, w)$} & &  & $\xi_\mathrm{hh}(x,z,M_1,M_2|\boldsymbol{\theta}_\mathrm{cosmo})$ & \\
   &  &  & FIT|PC|GP & $n_\mathrm{h}(z,M|\boldsymbol{\theta}_\mathrm{cosmo})$ & \\
 &  && NN & $P_\mathrm{h}^{(\mathrm{S})}(k,\mu|\boldsymbol{\theta}_\mathrm{cosmo},z,M_1,M_2)$ & \cite{Kobayashi20}
\end{tabular}
\begin{tablenotes}
\item[a] Near orthogonal design.
\item[b] Coulomb-like potential minimized. Each input variable then transformed to a normal or half-normal distribution.
\item[c] The sum of the distance from every point to the closest point for all two-dimensional projections is maximized.
\item[d] Axes of the hyperrectangle are rotated and rescaled to the eigenvectors and the eigenvalues of the posterior from previous CMB experiments.
\item[e] 4, 6 and the full 8D (sub)spaces each filled with points, glass configuration achieved by applying electrostatic repulsive force, antipodes discarded.
\item[f] Sample size gradually increased based on the uncertainties estimated by GP.
\end{tablenotes}
\end{threeparttable}
}
\end{table}

Emulation is now a popular approach for quick evaluation of various statistical properties of the large-scale structure of the universe. Table~\ref{tab:emulator} summarizes recent large-scale simulation campaigns for emulator building. The state-of-the-art emulators can predict not only the matter power spectrum, $P_\mathrm{m}(k)$ (or its projected version, $C_\ell$), the most fundamental quantity that characterizes the cosmological density field, but also many other quantities relevant to weak lensing observations
or those for \textit{galaxy} clustering observables. In Table 1, the quantities with superscript $\kappa$ are for the lensing convergence field\footnote{Here, convergence means the change in the observed size of an object caused by gravitational lensing effect. It is defined as the trace of the Jacobian between the unlensed and the lensed images.} in the ``Statistics'' column.
To be more precise, the cosmo-SLICS group studies the statistics of the aperture mass, which is closer to the actual observable.

For weak lensing, several non-Gaussian statistics are used to explore the information content beyond the power spectrum, including the bispectrum ($b_{\ell_1,\ell_2,\ell_3}$), the peak counts ($N_\mathrm{peak}$), the probability density function or the Minkowski functionals ($V_i$, $i=0,1,2$ for 2D lensing maps) (see \cite{Zurcher21} for a recent comparison among different statistics). For galaxy surveys, one needs to interpolate not only in the cosmological parameter space ($\boldsymbol{\theta}_\mathrm{cosmo}$), but also over the parameters specifying the properties of the galaxy sample of interest ($\boldsymbol{\theta}_\mathrm{galaxy}$), which makes the problem size even larger. To bypass this, one can rely on the so-called halo model approach (see \cite{Cooray02} for a review) while emulating the abundance and the clustering properties of dark matter halos as a function of their mass \cite{Nishimichi19,Kobayashi20}. One can also resort to perturbative bias expansion methods, such as the one in Lagrangian space \cite{Modi20}, for which various power spectra ($P_{ij}$ with $i$ and $j$ corresponding to different ``operators'') can be measured from simulation
outputs and then emulated~\cite{Kokron21,Hadzhiyska21,Zennaro21}. Some of the large simulation databases with different cosmological models, such as QUIJOTE~\cite{Paco20} and AbacusSummit~\cite{Maksimova21} are now publicly available.

We list large-scale cosmological simulation projects for emulators that interpolate summary statistics measured from simulations over the cosmological parameter space in Table~\ref{tab:emulator}. We note also that there are many other attempts to utilize emulators for different purposes, such as developing fast Boltzmann equation solvers or performing low-order perturbative calculations~\cite{Fendt07,Auld07,Auld08,Fendt09,Mootoovaloo20,Arico21b,DonaldMcCann22,DonaldMcCann22b,SpurioMancini22,Bonici22,DeRose22,Mootoovaloo22,Nygaard22,Eggemeier22}, to explore the galaxy-halo connection for fixed cosmology \cite{Kwan15}, to translate less costly, low-resolution simulations to mimic more expensive simulations. More specifically, the applications include incorporating baryonic effects to the dark-matter only simulations~\cite{Arico21}, predicting Lyman-$\alpha$ forest~\cite{Bird19,Pedersen21} or $21$-cm power spectra~\cite{Kern17,Schmit18}, extrapolating the predictions in $\Lambda$-Cold Dark Matter (CDM) cosmology
to alternative cosmological models~\cite{Giblin19}, or improving the spatial resolution of simulations~\cite{KodiRamanah20}. 
There are also promising approaches of mixed high- and low-resolution simulations as training data for summary-statistics emulators \cite{Ho22}. 
Emulators can be integrated as an essential part in Bayesian optimization by iteratively adding simulations or forward models in general to maximize the acquisition function, which quantifies the likelihood and uncertainties of the surrogate model~\cite{Leclercq18,PellejeroIbanez20,Boruah22,Neveux22}.

Recent studies successfully integrate emulators in cosmological parameter inference. 
A model prediction tool of Ref. \cite{Eifler11} based on the Coyote Universe emulator is used to calculate the cosmic-shear two-point correlation functions in the analysis of the SDSS weak lensing maps \cite{Huff14}.
Similarly, lensing peak counts and the power spectrum
are adopted in the analysis of CFHTLenS survey \cite{Liu15}, whereas the moments and Minkowski functionals are used in Ref. \cite{Petri15}.
Emulator-based peak counts are used for the analyses of DES Y1 \cite{HarnoisDeraps21} and DES Y3 data \cite{Zurcher22}.
Joint analyses 
have been performed using the angular clustering of SDSS galaxies and galaxy-galaxy lensing signal from Subaru HSC 
 \cite{Miyatake22}, and using large- and small-scale clustering of SDSS galaxies in redshift space \cite{Kobayashi22,Zhai22,Yuan22}. Modern statistical inference
often involves dense parameter sampling in a Monte Carlo manner, 
which can be performed only if the comparison with observed data
and various model predictions can be made fast enough.
In this sense emulators are indispensable tools for 
parameter inference in the big data era.

In the future, it would be ideal to adopt some kind of automated 
learning and parameter exploration.
There are a few important quantities that the next generation cosmology surveys
will be able to measure, such as the total mass of neutrinos, non-Gaussianities in the primordial fluctuations, spatial curvature and the dark energy
equation of state (e.g., \cite{DESI16,PFS14,Euclid13,Spergel15,Ivezic19,2014arXiv1412.4872D}).
The utilization of new statistical methods would significantly enhance the performance of these programs.

In ML applications, utilizing simulations as training data poses a current limitation in that there still remains a dearth of accurate and realistic mock observations of sufficient size. Recent studies  attempted to address this issue either by performing fast simulations at the expense of accuracy or by configuring smaller simulation volumes compared to the current or future observational programs, in order to generate a large number of mock realizations for proof-of-concept. However, even with a good prospect for accumulating large simulation datasets using future computational facilities, uncertainties surrounding the numerical implementation of non-gravitational effects persist. This is largely owing to both our limited understanding of the relevant astrophysical processes and the vast disparity in the physical length scale of galaxy formation and the observed volume. Therefore, addressing the challenges will require the development of efficient statistical approaches by introducing nuisance parameters and marginalizing them in the statistical inference, or by identification of robust summary statistics that are less susceptible to these uncertainties, to ensure reliable extraction of cosmological information.

\section{Concluding Remarks}
Brand-new telescopes will be in operation soon
and will explore the Universe with a broad range of wavelength. The amount of data delivered by the next generation observations is unprecedented.
It is urgently needed to develop efficient ML applications to process and analyze the sheer volume of data. This review article provides an overview of the ongoing effort in cosmology to make the best use of ML methods. 
Research in ML itself is developing rapidly,
where new approaches are proposed on a literally daily basis, not only for use in scientific research but also
for practical problems in our society.
Hopefully, astronomy can offer invaluable data sets for the ML/AI researchers to test and improve their methods by using multiplex data with a large volume
and of high quality.

Recent effort in observational cosmology largely focused on performing big-data analysis efficiently, and it is often left behind to address the overall reliability and interpretability of the machine-generated results. 
Developing explainable AI and gaining 
fundamental physics insights from data science approaches
will be important goals in the future \cite{Lundberg17,Mothilal20}.
Obviously, just bigger data from upcoming surveys do not automatically help improving theory predictions nor enhance modeling \cite{Hosni17}. 
Spurious correlations or even unphysical artifacts can be learned by sophisticated machines \cite{Laps19}.
It is already known that there are some certain limitations of currently popular generative models \cite{Konstantin22}.
In order to maximize the scientific returns
from future observations, another big leap would
be needed in ML/AI research so that the results -- often just numerics -- are understood in terms of physics and rigorous statistics.
Reasonable scientific implications can then be 
derived from ML approaches, which 
should ultimately lead to the development of a fully automated science research.

Admittedly, there are many topics that we have not been able to cover in this review. Most of the methods discussed here are successful examples adopted in astronomy and cosmology, but there is a much larger domain of ML applications other than basic science.
Completely new approaches may well be proposed from 
yet immature research areas. 
As we have emphasized repeatedly,
ML/AI is a hot research topic and is rapidly advancing in virtually all 
science research domains.
Frequent checks and updates 
will be necessary for the readers to catch up with the state-of-the-art. 
Thankfully, many of the papers introduced in this review offer their computer programs publicly available. The interested
readers should be able to find them easily in the respective papers,
and are encouraged to use for their own research or further code development.

\section*{Acknowledgments}
The authors acknowledge financial support by Japan Science and Technology Agency CREST JPMHCR1414 and by AIP Acceleration Research Grant Number JP20317829. This work was also supported in part by MEXT/JSPS KAKENHI Grant Number JP19H00677, JP20H05861, JP21H01081, and JP22K03634.

No new data were created or analysed in this study.

\section*{References}
\bibliography{MLCosmology}

\end{document}